\documentclass{article}
\usepackage{arxiv}

\usepackage[round]{natbib}

\usepackage{hyperref}
\usepackage{url}
\usepackage{amsthm}         % newtheorm
\usepackage{mathtools}      % coloneqq
\usepackage{relsize}        % mathsmaller
\usepackage{caption}
\usepackage{subcaption}
\usepackage{tikz}
\usepackage{algpseudocode} % algorithms
\usepackage{algorithm} % algorithms
\usepackage{setspace} % setstretch line spacing
\usepackage{dsfont} % /mathds for the indicator function
\usepackage{booktabs} % table separators
\usepackage{enumerate}
\usepackage{graphics} % resize tables

\newtheorem{definition}{Definition}
\newtheorem{example}{Example}

\newtheorem{lemma}{Lemma}
\newtheorem{proposition}{Proposition}
\newtheorem{remark}{Remark}

\usepackage{amsmath,amsfonts,bm}

\def\eqref#1{equation~\ref{#1}}
\def\1{\bm{1}}

\DeclareMathAlphabet{\mathsfit}{\encodingdefault}{\sfdefault}{m}{sl}
\SetMathAlphabet{\mathsfit}{bold}{\encodingdefault}{\sfdefault}{bx}{n}

\newcommand{\E}{\mathbb{E}}

\newcommand{\R}{\mathbb{R}}

\newcommand{\Var}{\mathrm{Var}}

\newcommand{\Cov}{\mathrm{Cov}}
\DeclareMathOperator*{\argmin}{arg\,min}

\newcommand{\Mu}{\operatorname{\mathbf{E}}} % Expectation
\newcommand{\Parents}{\mathbf{\operatorname{PA}}} % Parents
\newcommand{\Child}{\mathbf{\operatorname{CH}}} % Children 
\newcommand{\Varx}{\operatorname{Var}[\partial_\mathsmaller{U_X} \log p(U_X)]} % Var[s_X] bivariate model X -> Y
\newcommand{\Vary}{\operatorname{Var}[\partial_\mathsmaller{U_Y} \log p(U_Y)]} % Var[s_Y] bivariate model X -> Y
\newcommand{\Varf}{\operatorname{Var}[\partial_\mathsmaller{U_X}f(U_X)]}
\newcommand{\Covscore}{\Cov[\partial_\mathsmaller{U_X} \log p(U_X), \partial_\mathsmaller{U_X}f(U_X)\partial_\mathsmaller{U_Y} \log p(U_Y)]}
\newcommand{\h}{\mathbf{h}}
\newcommand{\x}{\mathbf{x}}

\DeclarePairedDelimiter{\abs}{\lvert}{\rvert}

\title{Shortcuts for causal discovery of nonlinear models by score matching}

\author{
Francesco Montagna$^1$\and
Lorenzo Rosasco$^{1, 2, 3}$\and
Nicoletta Noceti$^1$\and
Francesco Locatello$^4$
\affiliations
$^1$MaLGa, Università di Genova\\
$^2$MIT, CBMM\\
$^3$Istituto Italiano di Tecnologia (IIT)\\
$^4$Institute of Science and Technology Austria (ISTA)
}

\begin{document}

\maketitle

\begin{abstract}
\looseness-1The use of simulated data in the field of causal discovery is ubiquitous due to the scarcity of annotated real data. 
Recently, \citet{reisach21_beware} highlighted the emergence of patterns in simulated linear data, which displays increasing marginal variance in the casual direction. As an ablation in their experiments, \citet{montagna23_assumption} found that similar patterns may emerge in nonlinear models for the variance of the score vector $\nabla \log p_{\mathbf{X}}$, and introduced the ScoreSort algorithm. 
In this work, we formally define and characterize this \textit{score-sortability} pattern of nonlinear additive noise models.
We find that it defines a class of identifiable (bivariate) causal models overlapping with nonlinear additive noise models. We theoretically demonstrate the advantages of ScoreSort in terms of statistical efficiency compared to prior state-of-the-art score matching-based methods and empirically show the score-sortability of the most common synthetic benchmarks in the literature.
Our findings remark (1)  the lack of diversity in the data as an important limitation in the evaluation of nonlinear causal discovery approaches, (2) the importance of thoroughly testing different settings within a problem class, and (3) the importance of analyzing statistical properties in causal discovery, where research is often limited to defining identifiability conditions of the model. 
\end{abstract}

\section{Introduction}
\looseness-1The task of causal reasoning, framed as the ability to predict the effect of active interventions on a system, is central to virtually all scientific domains \citep{koller2009probabilistic, pearl09_causality, peters_2017}.
Frequently, manipulating the variables being investigated can be costly, challenging, or entirely unfeasible. This situation has spurred a growing interest in algorithms that identify causal relationships between measurables from purely observational data without needing the experimenter to actively intervene in the system. This inference problem is commonly known in the literature as \textit{causal discovery}. Summary information about causal relationships between the model variables is often represented in the form of directed acyclic graphs (DAGs) whose nodes are the variables of interest, and edges mark the existence of cause-effect relationships. Traditional causal discovery methods in the constraint and score-based literature are PC \citep{Spirtes2000} and GES \citep{chickering03_ges}: in the absence of restrictive assumptions on the causal model of the data generating process, these are limited to consistent inference of the Markov equivalence class of the causal graph, where some edges are left undirected, meaning that these methods are often not able to capture the asymmetry between cause and effect. Recently, it has been shown that restrictions on the class of functions generating effects from their causes ensure the \textit{identifiability} of the true DAG \citep{lingam_shimizu, hoyer08_anm, peters_2014_identifiability, zhang2009PNL}. These theoretical findings have drawn interest in defining algorithms for the inference of the causal graph underlying observational data \citep{shimizu11_dirlingam, peters14a_resit, buhlmann14_cam}. In the case of nonlinear additive noise models, a recent branch of the literature has investigated the connection between the score function $\nabla \log p(\mathbf{X})$ of the random vector $\mathbf{X}$ and its underlying causal model. In particular, the SCORE, DAS, and NoGAM algorithms \citep{rolland22_score, montagna23_das, montagna23_nogam} define conditions for the identification of the causal order and the edges of a causal graph by score matching estimation of the score function. 
 
Development and evaluation of causal discovery methodologies are significantly affected by the scarce availability of real data. As a consequence, researchers and practitioners tend to rely on synthetic data, which are the \textit{de facto} standard for the evaluation of novel methods.
Recently, \citet{reisach21_beware} and \citet{reisach2023_simple} brought to the attention the emergence of patterns in synthetic data generated according to causal models with linear functional mechanisms, which can be exploited to define simple heuristic algorithms achieving state-of-the-art performance on causal discovery. For example, they show that correct estimation of the causal order can be obtained by sorting variables by ascending order of their marginal variance in the case of observations generated through a linear causal model. Their work highlights the limitations of using synthetic data that strictly comply with some model specifications as the only resource for the evaluation of causal discovery methods, whereas it leaves as an open question under which condition assumptions compatible with the identified shortcuts in the data should be considered realistic.

\looseness-1In their recent paper, \citet{montagna23_assumption} conjectures the emergence of \textit{score-sortability}, a pattern in the variance of the score $\nabla \log p(\mathbf{X})$ of observations generated according to nonlinear additive noise models (ANMs), that tends to increase in the anti-causal direction. In our work,  we formally define and extensively investigate the score-sortability of nonlinear additive noise models from an empirical and theoretical perspective. 
We show that score-sortability defines a new class of identifiable causal models and that it can be exploited to attain state-of-the-art inference accuracy on synthetic datasets generated according to parameters commonly found in the literature, generally improving the statistical efficiency of SCORE, a causal discovery method based on the connection between the score function and the causal graph. Our contributions are summarized as follows:
\begin{itemize}
    \item We identify a pattern in the score function of data generated according to nonlinear ANM, showing that the variance of the score vector $\nabla \log p(\mathbf{X})$ increases in the anti-causal direction. This is the first work that focuses on the detailed study of patterns emerging in the setting of nonlinear data, whereas \citet{reisach21_beware} and \citet{reisach2023_simple} focus on linear models.
    \item We empirically show that the most common synthetic datasets for the evaluation of algorithms for nonlinear causal discovery are score-sortable. We regard this lack of diversity in the evaluation data as an important limitation in the literature.
    \item We demonstrate the state-of-the-art empirical performance of the \textit{ScoreSort} algorithm proposed in \citet{montagna23_assumption}, which finds the causal order of a graph by iterative identification of leaf nodes as the entries of the score vector where the variance is minimized.
    \item We define the necessary conditions for the identifiability of the causal order of a bivariate graph from observational data with ScoreSort. This defines a new class of identifiable causal models with partial overlap with the class of nonlinear ANM.
    \item \looseness-1We analyze the statistical properties of ScoreSort inference on score-sortable models, showing that it exhibits sample efficiency better than SCORE \citep{rolland22_score} under suitable assumptions. 
\end{itemize}

\section{Background and motivations}
\looseness-1In this section, we introduce the problem of causal discovery and the formalism of Structural Causal Models (SCMs). Then, we provide an overview of recent literature connecting the score function  (i.e. the gradient of the log-likelihood of the data) and the causal graph under the assumptions of the nonlinear additive noise model \citep{hoyer08_anm}.

\subsection{Problem definition}
A Structural Causal Model $\mathcal{M}$ is  defined by the tuple $(\mathbf{X}, \mathbf{U}, \mathcal{F}, p_\mathbf{U})$. This consists of the vector $\mathbf{X} \in \R^d$ of \textit{endogenous} random variables, vertices of the causal graph $\mathcal{G} = (\mathbf{X}, \mathcal{E})$ with the set of edges $\mathcal{E}$ that we want to identify. The vector of the \textit{exogenous} random disturbances $\mathbf{U} \in \R^d$, with the noise terms jointly distributed according to $p_\mathbf{U}$. The set of causal mechanisms $\mathcal{F} = (f_1, \ldots, f_d)$, deterministic maps assigning values to $X_1, \ldots, X_d$ respectively, given their causes and the corresponding error term $U_i$. Each variable $X_i$ is then defined by a structural equation:
\begin{equation}
    X_i \coloneqq f_i(\Parents_i, U_i), \hspace{1mm}\forall i=1,\ldots,d ,
    \label{eq:structural_equation}
\end{equation}
where $\Parents_i \subset \mathbf{X}$ is the set of parents of $X_i$ in the directed and acyclic causal graph $\mathcal{G}$, and denotes the set of direct causes of $X_i$. The recursive application of \eqref{eq:structural_equation} induces a joint distribution $p_\mathbf{X}$, such that the Markov factorization holds:
\begin{equation}
    p_\mathbf{X}(\mathbf{X}) = \prod_{i=1}^d p_i(X_i | \Parents_i).
    \label{eq:markov_factorization}
\end{equation}
Causal discovery aims to infer the the DAG $\mathcal{G}$ given a collection of $n$ observations drawn from the probability distribution $p_{\mathbf{X}}.$
From an algorithmic perspective, one common strategy is to separate the inference task into two steps, the first identifying the topological ordering between the nodes and the second finding the graph's edges admitted by such causal ordering. 

\paragraph{Topological order of a graph.} \looseness-1Given a directed acyclic graph $\mathcal{G} = (\mathbf{X}, \mathcal{E})$, one can define a partial ordering of the nodes $\pi = \{\pi_1, \ldots, \pi_d\}, \pi_i \in \{1, \ldots, d\}$, such that whenever we have $X_i \rightarrow X_j \in \mathcal{E}$, then $i \prec_{\pi} j$ ($j$ is a \textit{successor} of $i$ in the ordering $\pi$) \citep{koller2009probabilistic}. The permutation $\pi$ is known as the \textit{topological order} of $\mathcal{G}$, and allows to disambiguate the direction of the edges in the graph. This is crucial in the context of causal models, as knowledge of the topological order intrinsically distinguishes the cause from the effect between a pair of connected nodes. 

\paragraph{Identifiability of the causal graph.} Without further restrictions on the SCM of 
\eqref{eq:structural_equation}, it is not possible to infer the topological order of the causal graph from observational data, in which case we say that the model is not \textit{identifiable} \citep{peters_2017}. Instead, observations can inform about the Markov Equivalence Class (MEC) of the graph: given two DAGs, they belong to the same Markov equivalence class if they share the skeleton and the set of \textit{v-structures} (see Definition 6.24  in \citet{peters_2017}). The MEC can be represented as a CPDAG (Complete Partial DAG), where the direction of edges between two variables is often not specified. As a clarifying example, consider the pair of bivariate DAGs $X \rightarrow Y$ and $X \leftarrow Y$: given that they share the same skeleton, they belong to a unique MEC, represented by the undirected graph $X - Y$, where the asymmetry between cause and effect is not specified. In order to identify the topological order of a graph from observational data (i.e. in order to distinguish causes from effects), restrictions on the distribution of the noise terms $p_\mathbf{U}$ and on the class of functional mechanisms $\mathcal{F}$ are required. 

\subsection{Nonlinear Additive Noise Model}
\looseness-1Identifiability of the causal structure can be guaranteed under the assumptions of a nonlinear additive noise model \citep{hoyer08_anm, peters_2014_identifiability}, which defines the process generating causes from effects as a nonlinear deterministic function with additive noise terms. In particular, the ANM is defined  by \eqref{eq:structural_equation} when the following holds:
\begin{equation}
    X_i \coloneqq f_i(\Parents_i) + U_i, \hspace{1mm} \forall i=1,\ldots,d,
    \label{eq:anm}
\end{equation}
with $f_i$ nonlinear. Additional technical conditions on the class $\mathcal{F}$ of mechanisms and on the joint distribution of the noise terms are sufficient to ensure the identifiability of the model (see Condition 19 in \citet{peters_2014_identifiability}).

\subsection{The interplay between score matching and causal discovery}
Recent works in the literature have proven that it is possible to derive constraints on the gradient of the log-likelihood $\nabla \log p(\mathbf{X})$ (known as the score function) to identify both the topological order and the set of edges of a causal graph under the nonlinear additive noise model. \citet{rolland22_score}, \citet{montagna23_das}, and \citet{montagna23_nogam} exploits score matching \citep{hyvarinen_score_match} to define a  consistent estimator of the causal graph from observational data. The intuition is that, under identifiable conditions, it is possible to map a probability distribution $p_{\mathbf{X}}$ uniquely to the SCM generating the data. By application of the logarithm to the joint distribution $p_{\mathbf{X}}$, the product in the Markov factorization of \eqref{eq:markov_factorization} decomposes into a summation: $$\log p\textbf{}(\mathbf{X}) = \sum_{i=1}^d \log p_i(X_i | \Parents_i).$$ 
The score function is defined as the gradient of the log-likelihood. In the case of an additive noise model, for each node $X_i$ in the graph the corresponding entry in the score vector is $ s_i(\mathbf{X}) \coloneqq  \partial_{\mathsmaller{X_i}} \log p_{\mathbf{X}}(\mathbf{X})$, which equals to:
\begin{equation}
\begin{split}
       s_i(\mathbf{X}) &= \partial_{\mathsmaller{X_i}} \log p_i(X_i \mid \Parents_i) \\
       &+ \sum_{k \in \Child_i} \partial_{\mathsmaller{X_i}} \log p_k(X_k \mid \Parents_k),
\end{split}
   \label{eq:score_entry}
\end{equation}
where $\Child_i$ denotes the set of direct children of the node $X_i$. It is indeed important to notice that the summation takes place over the set of children: in the case of a \textit{leaf} $X_l$, i.e. a node with the set of children $\Child_l = \emptyset$, the corresponding component of the score $s_l(\mathbf{X})$ simplifies as follows:
\begin{equation}
    s_l(\mathbf{X}) \coloneqq \partial_{\mathsmaller{X_l}} \log p_{\mathbf{X}}(\mathbf{X}) =  \partial_{\mathsmaller{X_l}}\log p_l(X_l | \Parents_l).
    \label{eq:score_leaf}
\end{equation}
Notice that for nonlinear ANMs the summation over children vanishes if and only if the partial derivative of $\log p_{\mathbf{X}}(\mathbf{X})$ is relative to a leaf node. Intuitively, being able to capture this asymmetry between the entries of the score allows to infer the topological order of a causal graph from the data: \citet{rolland22_score} defines the conditions for the identifiability of the causal direction of nonlinear ANM  with 
Gaussian noise terms by deriving constraints on the score function, whereas \citet{montagna23_nogam} generalizes their results on arbitrary ANM without restrictions on the distribution of the noise random variables. The resulting SCORE and NoGAM algorithms (described in detail in Appendix \ref{app:score_nogam}) provide consistent estimators of the topological order via score matching inference of the gradient of the log-likelihood \citep{hyvarinen_score_match}. 

The score $\nabla \log p_{\mathbf{X}}(\mathbf{X})$ provides rich information about the causal model underlying the distribution, making the graph identifiable from pure observations. In the remainder of the paper, we show that even a simple heuristic to capture the asymmetry between the components of the score may be used to achieve state-of-the-art performance in causal discovery on data generated according to a nonlinear additive noise model.

\subsection{A simple baseline for causal order identification}
We have discussed how the structure of the score function can be used for the identification of the topological ordering of a nonlinear ANM. In particular, the problem of \textit{identifiability} of causal graphs amounts to finding asymmetries in the joint distribution of cause-effect pairs: being the $\nabla \log p_\mathbf{X}(\mathbf{X})$ a transformation of the distribution of the data, we expect the score vector to be informative about the direction of the causal relations. Having these considerations in mind, we observe that the variance of the score vector of an additive noise model cumulates in the anti-causal direction:  in the case of a bivariate graph $X \rightarrow Y$, we have indeed that $\Var[s_X(X, Y)] = \Var[\partial_{\mathsmaller{X}} \log p_X(X)] + \Var[\partial_{\mathsmaller{X}} \log p_Y(Y)] + C$, where $C$ is a covariance term, whereas $\Var[s_Y(X, Y)] = \Var[\partial_{\mathsmaller{Y}} \log p_Y(Y)]$. Comparing the two expressions, we get the intuition that the score of a leaf node can be characterized by a smaller variance with respect to the score of a node with children in the graph.
In the following example, we show a simple practical case in which the pattern in the variance of the score of a random variable generated according to a nonlinear causal model can be exploited to identify the topological order by a simple heuristic.

\begin{example}\label{example:score-sort}
Let $\mathbf{X} = (X_1, X_2, X_3)$ causally related according to a fully connected graph
$\mathcal{G}$, and assume the following simple SCM, such that closed-form computations are easy to perform:
\begin{equation}
    \begin{split}
        &X_1 \coloneqq U_1,\\
        &X_2 \coloneqq X_1^2 + U_2,\\
        &X_3 \coloneqq X_1^2 + X_2^2 + U_3,
    \end{split}
\end{equation}
where the noise terms are mutually independent random variables following a Gaussian distribution $\mathcal{N}(0, 1)$. 
The resulting entries of the score function are:
\begin{equation*}
    \begin{split}
        &s_1(\mathbf{X}) = U_1(2U_2 + 2U_3 -1)\\
        &s_2(\mathbf{X}) = U_2(2U_3 -1) +2U_1^2U_3\\
        &s_3(\mathbf{X}) = -U_3,
    \end{split}
\end{equation*}
and the vector of the variance of the score's components is $\Var[s(\mathbf{X})] = (9, 13, 1)$ (detailed computations can be found in Appendix \ref{app:example_computations}). Thus, we can identify the leaf node $X_3$ in the graph as the $\operatorname{argmin}_i\Var[s_i(\mathbf{X})]$. Given the topological order of the graph $\pi = (\pi_1, \pi_2, \pi_3)$, we find that the last element in the ordering is $\pi_3 = 3$. In order to find the complete topological ordering, we remove $X_3$ from the graph and iteratively repeat the procedure on the pruned graph $\tilde{\mathcal{G}}$ whose set of nodes is $\tilde{\mathbf{X}} \coloneqq (X_1, X_2)$. We obtain that the entries of the score function are $s_1(\tilde{\mathbf{X}}) = U_1(2U_2 -1)$ and $s_2(\tilde{\mathbf{X}}) = -U_2$, and the vector of the variance is equal to $\Var[s(\tilde{\mathbf{X}})] = (5, 1)$. As for the previous step, we find the index of the leaf node $X_2$ as the $\operatorname{argmin}_i\Var[s_i(\tilde{\mathbf{X}})]$. Thus we correctly conclude that the topological order of the graph $\mathcal{G}$ is $\pi = (1, 2, 3)$.
\end{example}

Next, our goal is to define formal conditions under which finding minimal variance in the score components can yield a topological order compatible with the causal graph of a nonlinear additive noise model.

\section{Score-sortability}
In the previous section, we discuss a pattern in the score of data generated according to nonlinear ANMs that is informative about the asymmetry in cause-effect relationships. In particular, Example \ref{example:score-sort} shows that the score entry of a leaf may be characterized by a smaller variance compared to the score associated with a node with children in the graph. In light of this consideration, we formalize a simple condition under which it is possible to identify leaf nodes of a causal graph from the variance of the score function. 
\begin{definition}[\textit{Score-identifiable leaf}]\label{def:score-sortability}
    Let $\mathbf{X} \in \R^d$ be a random vector defined by a set of structural equations as in \ref{eq:structural_equation}. Let $X_l$ be a leaf node of the causal graph associated with the SCM. We say that  $X_l$ is \textit{score-identifiable} if $l = \operatorname{argmin}_i \Var[s_i(\mathbf{X})]$.
\end{definition}
Example \ref{example:score-sort} illustrates the case of a causal graph whose leaf nodes are score-identifiable. 

\looseness-1Under the assumption of score-identifiable leaves, we can define an iterative procedure that finds the topological order associated with the set of causal variables $\mathbf{X} \in \R^d$. The details of this method are illustrated in the \textit{ScoreSort} Algorithm \ref{alg:score-sort-population} box, originally proposed in \citet{montagna23_assumption}. The idea is that at each iteration, a leaf node is identified as the $\operatorname{argmin}_i \Var[s_i(\mathbf{X})]$, and then it is removed from the graph. At the end of the iterating loop, the resulting output of the algorithm is a causal order $\pi^\textnormal{score}$ relative to the set of nodes $\mathbf{X}$.

Given a generic distribution $p_{\mathbf{X}}$ that is Markov with respect to the causal graph $\mathcal{G}$, it is not always the case that ScoreSort defines an ordering compatible with the DAG. Thus, we are interested in quantifying the agreement between $\pi^\textnormal{score}$ and the graph $\mathcal{G}$.

\begin{definition}[\textit{Score-sortability}]Let $\mathcal{G} = (\mathbf{X}, \mathcal{E})$ be a directed acyclic graph with set of nodes $\mathbf{X} \in \R^d$ generated according to a structural causal model $\mathcal{M}$, and with edges $\mathcal{E} = \{(i, j) : X_i \rightarrow X_j \}$. Moreover, let $\pi$ be the causal order output of Algorithm \ref{alg:score-sort-population}. We define the \textit{score-sortability} of $\mathcal{M}$ as follows:
    \begin{equation}
        \nu \coloneqq 1 - \frac{\sum_{(i,j) \in \mathcal{E}} \mathlarger{\mathds{1}}(j \prec_{\pi} i)}{\abs*{\mathcal{E}}} \in [0, 1],
    \end{equation}
    where $\mathlarger{\mathds{1}}$ is indicator function, $\abs*{\mathcal{E}}$ is the number of edges in the graph and $j \prec_{\pi} i$ denotes $i$ successor of $j$ in the ordering vector $\pi$.    
\end{definition}
Intuitively, the score-sortability counts the rate of edges in the ground truth DAG that are not admitted by the ordering $\pi$ found with Algorithm \ref{alg:score-sort-population}: the rate is then subtracted to $1$, such that $\nu = 1$ when $\pi$ is correct with respect to the graph. For example, the score-sortability of the model in Example $\ref{example:score-sort}$ is $\nu=1$, which corresponds to an identifiable causal graph. A score-sortability value $\nu = 0.5$ denotes that the output of ScoreSort is equivalent to the expected accuracy of a random ordering. Next, we show that score-sortability defines a new class of identifiable causal models.
\begin{algorithm}
\caption{\looseness-1ScoreSort (finite sample estimation in the comments), adapted from \citet{montagna23_assumption}}\label{alg:score-sort-population}
\setstretch{1.4}
\begin{algorithmic}
\State $\mathbf{X} \in R^d$, $\mathbf{X} \hspace{.8mm} \mathsmaller{\sim} \hspace{.8mm} p_{\mathbf{X}}$ \hspace{.7em}\textcolor{blue}{\texttt{//} $\mathit{X} \in \R^{n \times d}$}

\State $\pi \leftarrow []$

\State $\textnormal{nodes} \leftarrow [1, \ldots, d]$\

\For{$i = 1, \ldots, d$}
    
    \State $s(\mathbf{X}) \leftarrow \nabla \log p_\mathbf{X}(\mathbf{X})$ \hspace{.5em} \textcolor{blue}{\texttt{//} \texttt{score-matching$(\mathit{X})$}}
    
    \State $\lambda \leftarrow \operatorname{argmin} \Var[s(\mathbf{X})]$ \hspace{.5em} \textcolor{blue}{\texttt{//} $ \operatorname{argmin}\hat{\Var}[\hat{s}(\mathbf{X})]$}

    \State $l \leftarrow \textnormal{nodes}[\lambda]$

    \State $\pi \leftarrow [l, \pi]$

    \State  Remove $\lambda$-th entry from $\mathbf{X}$ \hspace{.5em}\textcolor{blue}{\texttt{//} Remove $\mathit{X}[:, \lambda] $}

    \State Remove $l$ from nodes
    
\EndFor

\State \Return $\pi$

\end{algorithmic}
\end{algorithm}

\subsection{ScoreSort identifiability of the bivariate model} We propose sufficient conditions for the identifiability of a  bivariate additive noise model of the form $X \coloneqq U_X$, $Y \coloneqq f(U_X) + U_Y$, corresponding to the graph $X \rightarrow Y$. It is immediate to see that the model is identifiable by ScoreSort if and only if $\Var[s_X(X, Y)] > \Var[s_Y(X, Y)]$. From  \eqref{eq:score_entry}, we can derive the variance of the score components: 
\begin{gather}
    \begin{split}
        \Var[s_X(X, Y)] &= \Varf \Vary  \\
        &+ \Varx + 2C
    \end{split} \label{eq:bivariate_score_x}\\[.5em]
    \hspace{-6.5em} \Var[s_Y(X, Y)] =  \Vary , \label{eq:bivariate_score_y}
\end{gather}
where, with an abuse of notation, the different probability distributions $p$ are discerned by their respective arguments. As a shortcut notation, we also define $C \coloneqq \Covscore$.
\begin{proposition}\label{prop:identifiability}
    Let $X \rightarrow Y$ be the graph associated with a causal model with structural equations $X \coloneqq U_X$, $Y \coloneqq f(U_X) + U_Y$. Then:
    \begin{equation*}
        \begin{split}
\nu = 1 \Longleftrightarrow &\Varf > 1 - \frac{\Varx}{\Vary} \\
&-\frac{2C}{\Vary}.
        \end{split}
    \end{equation*}
\end{proposition}
According to Proposition \ref{prop:identifiability}, the bivariate additive noise model is score-sortable and hence identifiable when the variance of $\partial_\mathsmaller{U_X}f(U_X)$ is \textit{sufficiently} large. Note that this is never the case under the hypothesis of linear causal mechanisms. A proof for Proposition \ref{prop:identifiability} is provided in Appendix \ref{app:prop_proof}.

\begin{remark}\label{remark:prop1}
    \looseness-1Score-sortability is not limited to the case of nonlinear additive noise models, as the structure of the score function of \eqref{eq:score_entry} holds for generic causal models that satisfy the Markov factorization in \eqref{eq:markov_factorization}. Hence, score-sortable models define a new class of identifiable SCMs, which includes additive noise models restricted to the case satisfying the condition of Proposition \ref{prop:identifiability}.
\end{remark}

\subsection{Score-sortability of ANM datasets}
\looseness-1In the case when we can not access the distribution $p_{\mathbf{X}}$, but only a finite set of $n$ observations $\mathit{X}\in \R^{n \times d}$, we can not assess the score-sortability of the model directly. In practice, we can exploit the \textit{ScoreSort} algorithm in the finite samples regime (refer to the Algorithm \ref{alg:score-sort-population} box): instead of computing the score function $s(\mathbf{X})$ directly from the distribution of the data, this is inferred via score matching by the Stein gradient estimator \citep{stein_gradient} (see Appendix \ref{app:stein_gradient}), which provides a consistent estimator $\hat{s}(\mathbf{X})$ of the score. Thus, the output of the ScoreSort algorithm is a consistent estimator of the score-sortability of the causal model of interest. In the next section, we discuss the statistical efficiency of ScoreSort in comparison to that of the SCORE algorithm.

\subsection{Comparing ScoreSort and SCORE statistical efficiency}
\looseness-1In practice, the main difference between SCORE and ScoreSort decision rules for leaf node identification is that SCORE relies on the estimation of the Hessian matrix $\nabla^2 \log p(\mathbf{X})$, whereas ScoreSort is based on the inspection of the first order partial derivatives in the gradient of the log-likelihood. The key point for the comparison of the two algorithms' statistical efficiency is that the Hessian estimator defined in SCORE is found by minimizing the error of a regression problem, which requires access to the score vector $\nabla \log p(\mathbf{X})$: given that this is generally unknown, it is replaced by its score matching estimate  $\nabla \widehat{\log p(\mathbf{X})}$. Intuitively, errors in $\nabla \widehat{\log p(\mathbf{X})}$ due to the finiteness of the sample propagate in the values inferred for the Hessian matrix $\nabla^2 \log p(\mathbf{X})$. In what follows, we denote $\widehat{\partial_{x_i} \log p(\mathbf{x})}$ as the score matching estimator of the score entry $s_i(\mathbf{x})$ as defined in the ScoreSort algorithm, and $\widehat{\partial^2_{x_i} \log p(\mathbf{x})}$ as the estimator of the second order partial derivative of the log-likelihood, as defined in SCORE. 

\begin{proposition}\label{prop:stat_efficiency} Let $X \in \R^{n \times d}$ be a sample generated according to a structural causal model as defined in \eqref{eq:structural_equation}. 
Let $\delta_i^{(k)} \coloneqq \abs{\partial_{x_i} \log p(\mathbf{x}^{(k)}) - \widehat{\partial_{x_i} \log p(\mathbf{x}^{(k)})}}$, where $\mathbf{x}^{(k)} \in \R$ is a row of the matrix $X$. Let also $\epsilon_i^{(k)} \coloneqq \abs{\partial^2_{x_i} \log p(\mathbf{x}^{(k)}) -  \widehat{\partial^2_{x_i} \log p(\mathbf{x}^{(k)})}}$. Assume that if $\widehat{\partial_{x_i} \log p(\mathbf{x}^{(k)})} = \partial_{x_i} \log p(\mathbf{x}^{(k)})$, then $\widehat{\partial^2_{x_i} \log p(\mathbf{x}^{(k)})} = \partial^2_{x_i} \log p(\mathbf{x}^{(k)})$. Then, 
$$\epsilon_i^{(k)} = \delta_i^{(k)}\abs*{\partial_{x_i} \log p(\mathbf{x}^{(k)}) + \widehat{\partial_{x_i} \log p(\mathbf{x}^{(k)})}}.$$
\end{proposition}

\looseness-1Intuitively, the statistical error of first-order partial derivatives propagates in the second-order estimators: as  $\abs*{\partial_{x_i} \log p(\mathbf{x}^{(k)}) + \widehat{\partial_{x_i} \log p(\mathbf{x}^{(k)})}}$ gets larger, we expect ScoreSort to display statistical efficiency better than SCORE in the inference of the causal graph. (Proof in Appendix \ref{app:stat_efficiency}.)

\begin{remark}
    \looseness-1 To simplify the analysis, Proposition \ref{prop:stat_efficiency} assumes that if the score $\partial_{x_i} \log p(\mathbf{x}^{(k)})$ is exactly known,  then the regression error for the second-order estimator also vanishes. In practice, this is not guaranteed, and SCORE may have even larger errors.
\end{remark}

\begin{remark}
Proposition \ref{prop:stat_efficiency} highlights the connection between the assumptions of identifiability of a causal model and the statistical error in the inference of the graph. In particular, in score-sortable settings, ScoreSort represents a baseline with statistical efficiency better than SCORE for sufficiently large values of  $\abs*{\partial_{x_i} \log p(\mathbf{x}^{(k)}) + \widehat{\partial_{x_i} \log p(\mathbf{x}^{(k)})}}$.
\end{remark}

\section{Experimental results}\label{sec:experiments}
\looseness-1In this section, we investigate ScoreSort's empirical performance, as well as the score-sortability of real and synthetic data commonly used for the evaluation of nonlinear causal discovery algorithms.

\paragraph{Methods.} \looseness-1We compare ScoreSort accuracy with order-based methods regarded as state-of-the-art for inference on additive noise models, namely SCORE, NoGAM, CAM \citep{buhlmann14_cam}, and RESIT \citep{peters14a_resit} algorithms\footnote{We consider the DoDiscover implementation of SCORE, NoGAM, CAM \citep{Li_Dodiscover_Causal_discovery}, and a custom implementation of RESIT based on the \href{https://github.com/cdt15/lingam/blob/master/lingam/resit.py}{LiNGAM repository.}} (see Appendix \ref{app:methods}).

\paragraph{Metrics.} \looseness-1In order to evaluate the score-sortability of a causal model, we use the $\textnormal{FNR-}\hat{\pi}$ accuracy introduced in \citet{montagna23_assumption}, that measures the false negative rate against the ground truth of the unique fully connected graph compatible with the topological order $\hat{\pi}$. Given a sorting $\hat{\pi}$ inferred from a causal discovery method, the $\textnormal{FNR-}\hat{\pi}$ is defined as the false negative rate of the DAG with edges $\mathcal{E}_{\hat{\pi}} = \{X_{\hat{\pi}_i} \rightarrow X_{\hat{\pi}_j}: \hat{\pi}_i \prec_{\hat{\pi}} \hat{\pi}_j \hspace{1mm}, \forall i,j=1,\ldots,d \}$. In the case of a fully connected graph, a false negative corresponds to an edge with the direction reversed with respect to the target. If $\hat{\pi}$ is correct with respect to the ground truth graph, then $\textnormal{FNR-}\hat{\pi} = 0$. When $\hat{\pi}$ is the order inferred with the ScoreSort algorithm, $\textnormal{FNR-}\hat{\pi}$ evaluates the score-sortability of the causal graph, which can be simply found as $\nu = 1 - \textnormal{FNR-}\hat{\pi}$. Additionally, we use the Structural Hamming Distance (SHD), counting the number of missing and reversed edges in the prediction. The SHD records are reported in Section \ref{app:er_shd} of the appendix, as our main goal is the analysis of the score-sortability of the data.

\subsection{Score-sortability of synthetic data}
The most common strategy for the generation of synthetic causal graphs consists of stochastic sampling of an acyclic graph, randomly generating the causal mechanisms either as a Gaussian process (\textit{GP data}) or via a transformation defined by a neural network (\textit{NN data}) (a thorough list of references where this data simulation setting is employed can be found in Appendix \ref{app:references}). The standard practice is to generate the causal graphs with the Erd\"{o}s-Renyi \citep{Erdos:1960} and the Scale-free models \citep{Barabasi99emergenceScaling} (experiments on Scale-free networks are reported in Appendix \ref{app:scale-free-exp}). In our experiments, we consider datasets of $1000$ samples of sparse and dense graphs with $\{5, 10, 20, 50\}$ nodes 
(Appendix \ref{app:synthetic_data} for details on the data generation).
\begin{figure*}
    \centering
    \begin{subfigure}[b]{1\textwidth}
        \centering
        \includegraphics[width=0.57\textwidth]{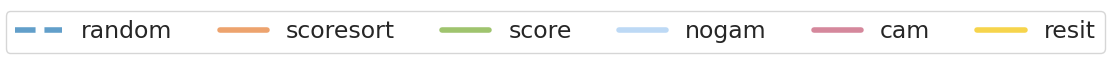}
    \end{subfigure}
     \begin{subfigure}[b]{1\textwidth}
        \centering        
    \includegraphics[width=0.9\textwidth]{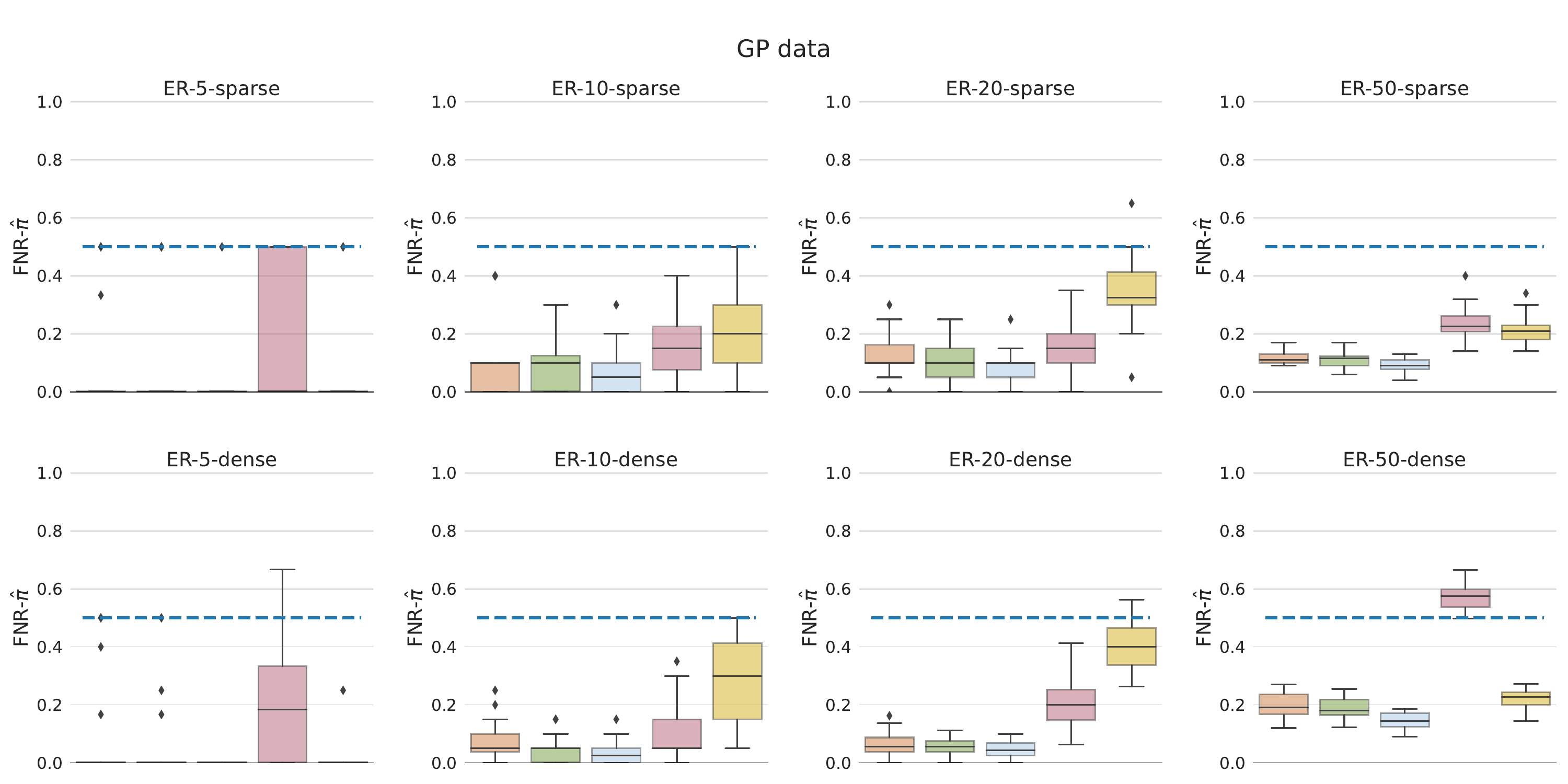}
    \end{subfigure}
    \caption{\footnotesize{Experimental results on dense and sparse ER with number of nodes in the set $\{5, 10, 20, 50\}$ and GP causal mechanisms. The graphs show the $\textnormal{FNR-}\hat{\pi}$ accuracy (the lower, the better) of the order estimates, with the boxplots evaluated over $20$ random seeds. Score-sortability is defined as $\nu = 1 - \textnormal{FNR-}\hat{\pi}$ achieved by ScoreSort, such that low values of $\textnormal{FNR-}\hat{\pi}$ denotes high score-sortability of the model. The dashed blue line is the expected accuracy of random ordering.}}
    \label{fig:exp_GP}
\end{figure*}
\paragraph{GP data experiments.} \looseness-1Figure \ref{fig:exp_GP} illustrates the empirical results on additive noise model synthetic data with Erd\"{o}s-Renyi graphs and causal mechanisms sampled from a Gaussian process. We observe that the ScoreSort algorithm performance is comparable to that of SCORE and NoGAM for all combinations of density and graph size while being comparable to or significantly better than CAM and RESIT. Overall, we conclude that \textit{GP data} are score-sortable, given that ScoreSort achieves $1 - \textnormal{FNR-}{\hat{\pi}}$ (estimate of the data score-sortability) with median in the range $[0.8, 1.0]$.

\begin{figure*}
    \centering
    \begin{subfigure}[b]{1\textwidth}
        \centering
        \includegraphics[width=0.57\textwidth]{Figures/legend_scoresort.png}
    \end{subfigure}
     \begin{subfigure}[b]{1\textwidth}
        \centering        
    \includegraphics[width=0.93\textwidth]{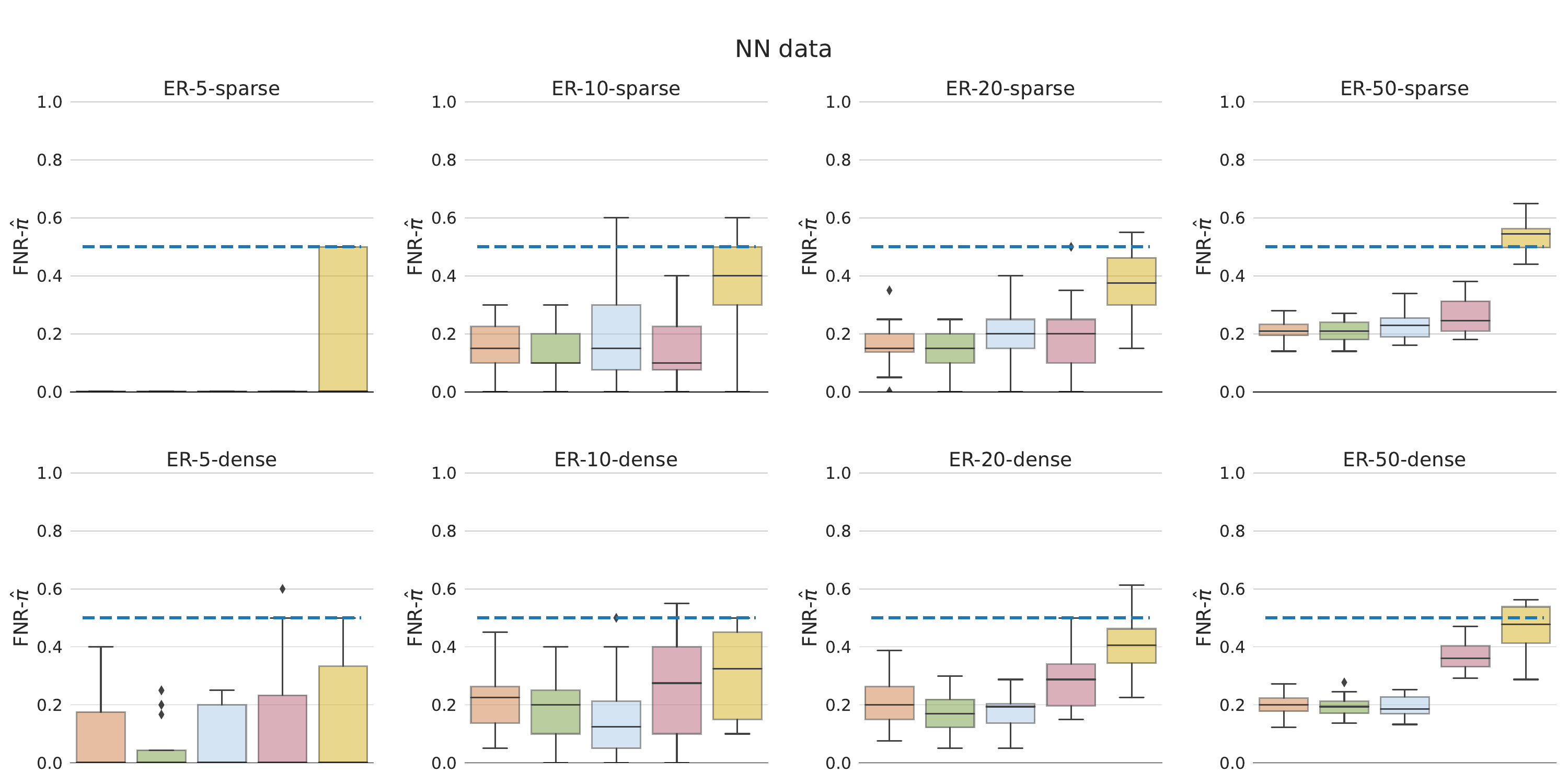}
    \end{subfigure}
    \caption{\footnotesize{Experimental results on dense and sparse ER with number of nodes in the set $\{5, 10, 20, 50\}$ and NN causal mechanisms. The graphs show the $\textnormal{FNR-}\hat{\pi}$ accuracy (the lower, the better) of the order estimates, with the boxplots evaluated over $20$ random seeds. Score-sortability is defined as $\nu = 1 - \textnormal{FNR-}\hat{\pi}$ achieved by ScoreSort, such that low values of $\textnormal{FNR-}\hat{\pi}$ denotes high score-sortability of the model. The dashed blue line is the expected accuracy of random ordering.}}
    \label{fig:exp_NN}
\end{figure*}

\paragraph{NN data experiments.} \looseness-1Figure \ref{fig:exp_NN} shows the empirical results on data simulated from nonlinear ANMs and Erd\"{o}s-Renyi graphs, and mechanisms parametrized by neural networks. 
We see that ScoreSort generally infers the topological order with accuracy consistently better than random, comparable to that of SCORE and NoGAM. We conclude that \textit{NN data} are generally characterized by high score-sortability, given that the median values of the estimated score-sortability are in the range $[0.8, 1.0]$.

\paragraph{Implications.} \looseness-1Our experiments demonstrate the score-sortability of the most popular simulated data for the evaluation of nonlinear causal discovery methods, showcasing limitations in the diversity of these common benchmarks. Instead, we advocate for (1) testing the score-sortability of any proposed future benchmark and (2) extending the evaluation of causal discovery methods beyond score-sortable datasets.
% NN plot

\subsection{Score-sortability of real data}
\looseness-1In this section, we discuss the score-sortability of real and semi-synthetic data. We consider a biological dataset of gene expression records with 17 edges and 853 observations, known as Sachs data \citep{sachs_2005} (a common benchmark in the causal discovery literature). Additionally, we experiment on 20 distinct semi-synthetic datasets sampled from SynTReN generator of realistic gene expression records \citep{syntren_2006}, consisting of $1000$ samples from a casual graph with $20$ nodes and variable number of edges from $19$ to $34$. In Table \ref{tab:exp_real}, we observe that both score-sortability and the benchmarked methods' performance decrease when compared to experiments on simulated data. Given that real data allow no control over the process generating the observations, these results may be explained by the fact that the model underlying the samples does not comply with the nonlinear ANM and the score-sortable model's hypothesis.

\begin{table}[]
\resizebox{\columnwidth}{!}{
    \setlength{\tabcolsep}{5.pt}
    \centering
    \begin{tabular}{lccccc}
         & ScoreSort & SCORE & 
         NoGAM & CAM & RESIT \\
         \toprule
         {SACHS} & $\mathbf{0.47}$ & $0.47$ & $0.47$ & $0.47$ & $\mathbf{0.35}$  \\
         SynTReN & $\mathbf{0.52 \pm 0.1}$ & $0.55 \pm 0.13$ & $0.54 \pm 0.11$ & $\mathbf{0.5 \pm 0.16}$ & $0.64 \pm 0.12$ \\
         \bottomrule \\
    \end{tabular}
}
    \caption{$\textnormal{FNR-}\hat{\pi}$ (the lower, the better) on the SynTReN and Sachs datasets. For SynTReN data, we report mean and standard deviation on $20$ random seeds.}
    \label{tab:exp_real}
\end{table}

\subsection{Discussion}
\looseness-1Our experiments on simulated environments show that the most common synthetic benchmarks for nonlinear causal discovery, with mechanisms sampled from Gaussian processes and random neural networks, consist of \textit{score-sortable} causal models, such that the variance of the score vector can be used to identify the topological order with state-of-the-art performance. This implies that when relying exclusively on \textit{GP} and \textit{NN data} generated according to our parameters, the experiments probe the inference ability of causal discovery methods in the restricted class of score-sortable models. In our Proposition \ref{prop:identifiability} we show that ANMs may not satisfy score-sortability, while the assumptions for score-sortability do not directly imply the additive noise model (Remark \ref{remark:prop1}): then, empirical evaluation bounded to score-sortable scenarios provides biased information on the performance of methods requiring the ANM hypothesis, restricted to the subclass of inference problems where the simple ScoreSort baseline represents the state of the art. This lack of diversity in the data posits a fundamental limitation in the evaluation of causal discovery approaches for the nonlinear additive noise model: we advise that (1) any future benchmark should assess the score-sortability of the data and (2) that meaningful evaluation should not be limited to score-sortable scenarios.

\section{Conclusion}
We characterize the score-sortability pattern emerging in data generated by nonlinear causal models, where the variance of the components of the score $\nabla \log p_\mathbf{X}(\mathbf{X})$ increases in the anti-causal direction. This property of the data can be exploited for the identification of the causal order: we show that score-sortable causal models are accurately inferred by ScoreSort, which generally improves the statistical efficiency of the SCORE algorithm for ANMs. Our contribution extends to the nonlinear setting the discussion on patterns arising in simulated data presented in \citet{reisach21_beware} and \citet{reisach2023_simple}. As one of our key findings, we show that the most common synthetic benchmarks for the evaluation of methods for causal discovery on additive noise models are all characterized by high values of score-sortability. Given that the set of score-sortable models only partially overlaps with ANMs, this implies that the most common evaluation strategies in the literature only provide a biased view of the algorithms' performance, limited to a subclass of the casual models satisfying the required ANM assumptions. We leave as future work the study of alternative patterns emerging in nonlinear scenarios beyond the restricted case of additive noise models, as well as the characterization of their plausibility in real-world applications.

\bibliography{bibliography}

\onecolumn
\appendix

\section{Score matching-based causal discovery}\label{app:score_nogam}
In this section, we present an overview of the ideas behind the SCORE and NoGAM algorithms, that exploit score matching estimation of the gradient of the log-likelihood to infer the topological ordering of nonlinear additive noise models.

\paragraph{SCORE.} \citet{rolland22_score} defines a formal criterion for the identification of the causal order of a graph underlying an additive noise model with Gaussian distribution of the noise terms. The intuition is that, under these assumptions, the second order partial derivative $\partial_{\mathsmaller{X_l}} s_l(\mathbf{X})$ is a constant if and only if $X_l$ is a leaf.
\begin{lemma}[Lemma 1 of \citet{rolland22_score}]
% \label{lem:score}
Let $\mathbf{X}$ be a random vector generated according to an identifiable ANM with exogenous noise terms $U_i \sim \mathcal{N}(0, \sigma_i^2)$, and let $X_i \in \mathbf{X}$. Then 
\begin{equation}
    \Var \left[ \partial_{\mathsmaller{X_i}}s_i(\mathbf{X}) \right] = 0 \Longleftrightarrow X_i \textnormal{ is a leaf, } \: \forall i = 1, \ldots, d.
    % \label{eq:var_lemma_rolland}
\end{equation}
\end{lemma}
The authors define the SCORE algorithm for the inference of the topological order, given a dataset of i.i.d. observations $X \in \R^{n \times d}$: first, SCORE estimates the diagonal elements of the Jacobian matrix of the score $J_\mathbf{s}$ via score matching (using an extension of the Stein gradient estimator proposed by \citet{stein_gradient}, discussed in its details in Appendix \ref{app:stein_hessian}). Then, it identifies a leaf in the graph as the $\operatorname{argmin}_i \Var[\partial_{\mathsmaller{X_i}}s(\mathbf{X})]$, which is removed from the graph and assigned a position in the order vector. By iteratively repeating this two-steps procedure up to the source nodes, all variables in $\mathbf{X}$ eventually are assigned a position in the causal ordering.

\paragraph{NoGAM.} \citet{montagna23_nogam} exploits the score function to define a formal criterion for the identification of leaf nodes in a graph induced by an additive noise model without restrictions on the distribution of the noise terms. After some manipulations, it can be shown that the score entry of a leaf $X_l$ defined in \eqref{eq:score_leaf} satisfies
\begin{equation}
    s_l(\mathbf{X}) = \partial_{\mathsmaller{U_l}} \log p_l(U_l),
    % \label{eq:score_leaf_noise}
\end{equation}
\looseness-1such that observations of the pair $(U_l, s_l(\mathbf{X}))$ can be used to learn a predictor of the score entry. For an additive noise model, the authors show that the noise term of a leaf is equal to the residual defined as:
\begin{equation}
    R_l \coloneqq X_l - \Mu\left[X_l \mid \mathbf{X}\setminus X_l\right].
    \label{eq:app_nogam_residuals_leaf}
\end{equation} 
Then, it is possible to find a consistent approximator of the score entry of a leaf node using $R_l$ as the only predictor. 
\begin{lemma}[Lemma 1 of \citet{montagna23_nogam}]
\label{lem:nogam}
Let $\mathbf{X}$ be a random vector generated according to an identifiable ANM, and let $X_i \in \mathbf{X}$. Then $$\Mu\left[\left(\mathbf{E}\left[s_i(\mathbf{X}) \mid R_i\right] - s_i(\mathbf{X})\right)^2\right] = 0 \Longleftrightarrow X_i \textnormal{ is a leaf.}$$
\end{lemma}
\looseness-1Similarly to SCORE, NoGAM algorithm uses score matching estimation to define a procedure for the inference of the topological order by iterative identification of leaf nodes, which are found as the $\operatorname{argmin}_i \Mu\left[\left(\mathbf{E}\left[s_i(\mathbf{X}) \mid R_i\right] - s_i(\mathbf{X})\right)^2\right]$. The residuals $R_i, i=1, \ldots, d,$ can be estimated by any regression algorithm.

Once the order is found, both SCORE and NoGAM algorithms select the edges by pruning the fully connected graph compatible with the topological order. This procedure is called \textit{CAM-pruning} and is described in detail in the next section.

\section{Other methods}\label{app:methods}
Now, we provide details on CAM and RESIT algorithms benchmarked in the experimental section \ref{sec:experiments}.

\subsection{CAM}\label{app:cam}
CAM algorithm \cite{buhlmann14_cam} infers a causal graph from data generated by an additive Gaussian noise model. 
First, it infers the topological ordering by finding the permutation of the graph nodes corresponding to the fully connected graph that maximizes the log-likelihood of the data.
After inference of the topological ordering, a pruning step is done by variable selection with regression. In particular, for each variable $X_j$ CAM fits a generalized additive model using as covariates all the predecessor of $X_j$ in the ordering, and performs hypothesis testing to select relevant parent variables. This is known as the \textit{CAM-pruning} algorithm. For graphs with size strictly larger than $20$ nodes,  the authors of CAM propose an additional preliminary edge selection step, known as Preliminary Neighbours Search (PNS): given an order $\pi$, variable selection is performed by fitting for each $j = 1, \ldots, d$ an additive model of $X_j$ versus all the other variables $\{X_i: X_j \succ X_i \textnormal{ in } \pi\}$, and choosing the $K$ most important predictor variables as possible parents of $X_j$. This preliminary search step allows scaling CAM pruning to graphs of large dimensions. In our experiments, CAM-pruning is implemented with the preliminary neighbors search only for graphs of size $50$, with $K=20$.

\subsection{RESIT}
In RESIT (regression with subsequent independence test) \cite{peters_2014_identifiability} the authors exploit the independence of the noise terms under causal sufficiency to identify the topological order of the graph. For each variable $X_i$, they define the residuals $R_i = X_i - \Mu\left[X_i \mid \mathbf{X} \setminus \{X_i\} \right]$, such that for a leaf node $X_l$ it holds that $R_l = U_l - \Mu[U_l]$. The method is based on the property that under causal sufficiency, the noise variables are independent of all the preceding variables: after estimating the residuals from the data, it identifies a leaf in the graph by finding the residual $R_l$ that is unconditionally independent of any node $X_i, \forall i\neq l$ in the graph. Once an order is given, they select a subset of the edges admitted by the fully connected graph encoding of the ordering. We implement this final step with CAM-pruning.

\section{Example 1}\label{app:example_computations}
In this section, we provide detailed computations of the variance of the score vector relative to  Example \ref{example:score-sort}. Given the  structural causal model
\begin{equation}\label{eq:example_scm}
    \begin{split}
        &X_1 \coloneqq U_1,\\
        &X_2 \coloneqq X_1^2 + U_2,\\
        &X_3 \coloneqq X_1^2 + X_2^2 + U_3,
    \end{split}
\end{equation}
under the assumption of mutually independent noise terms with Gaussian distribution $\mathcal{N}(0, 1)$, according to \eqref{eq:score_entry} the analytic form of the score components is:
\begin{equation*}
    -(X_i - f_i(\Parents_i)) + \sum_{k \in \Child_i}\partial_\mathsmaller{X_i}f_k(\Parents_k)(X_k - f_k(\Parents_k)).
\end{equation*}
Thus, the score entries for the model of \eqref{eq:example_scm} are:
\begin{equation*}
    \begin{split}
        &s_1(\mathbf{X}) = U_1(2U_2 + 2U_3 -1)\\
        &s_2(\mathbf{X}) = U_2(2U_3 -1) +2U_1^2U_3\\
        &s_3(\mathbf{X}) = -U_3.
    \end{split}
\end{equation*}
Now, we proceed with the computation of the marginal variance of the vector components.
The variance of $s_1(\mathbf{X})$ is given by:
\begin{equation}\label{eq:var_s1_example}
    \begin{split}
        \Var[s_1(\mathbf{X})] &= 4\Var[U_1U_2] + 4\Var[U_1U_3] + \Var[U_1] \\
        &= 4\Var[U_1]\Var[U_2] + 4\Var[U_1]\Var[U_3] + \Var[U_1] \\
        &= 4 + 4 + 1 = 9.
    \end{split}
\end{equation}
 It is easy to prove that in \eqref{eq:var_s1_example} the covariance terms given by the sum of random variables vanish. 
For the score entry $s_2(\mathbf{X})$, we get:
\begin{equation*}
    \begin{split}
        \Var[s_2(\mathbf{X})] &= \Var[U_2(2U_3 -1) +2U_1^2U_3] \\
        &= \Var[- U_2 + 2U_2U_3 + 2U_1^2U_3] \\
        &= \Var[U_2] + 4\Var[U_2U_3] + 4\Var[U_1^2U_3] \\
        &= 1 + 4\Var[U_2]\Var[U_3] + 4\Var[U_1^2]\Var[U_3] \\
        &= 1 + 4 + 8 = 13.
    \end{split}
    \label{eq:var_s2_example}
\end{equation*}
Similarly to the previous case, trivial computations show vanishing covariance. Finally, we can immediately conclude that $\Var[s_3(\mathbf{X})] = \Var[U_3] = 1$, hence the vector of marginal variances of the score is $(9, 13, 1)$. Thus, we correctly 

Next, we consider the calculation of the marginal variance of the score of the pruned graph $\tilde{\mathcal{G}}$ whose set of nodes is $\tilde{\mathbf{X}} \coloneqq (X_1, X_2)$. The score components are given by:
\begin{equation*}
    \begin{split}
        &s_1(\tilde{\mathbf{X}}) = U_1(2U_2 -1)\\
        &s_2(\tilde{\mathbf{X}}) = -U_2.
    \end{split}
\end{equation*}
The marginal variance of the first component is:
\begin{equation*}
    \begin{split}
        \Var[s_1(\tilde(\mathbf{X})] &= \Var[U_1(2U_2 -1)] \\
        &= \Var[2U_1U_2] + \Var[U_1] \\
        &= 4\Var[U_1]\Var[U_2] + 1 \\
        &= 4 + 1 = 5.
    \end{split}
\end{equation*}
\looseness-1Finally, we have $\Var[s_2(\tilde(\mathbf{X}))] = \Var[U_2] = 1$, such that the vector of marginal variance of the score of $\tilde{\mathbf{X}}$ is (5, 1).

\section{Synthetic data}
\begin{table}
  \centering
  \begin{tabular}{lllll}
    \toprule
     & $5$ nodes & $10$ nodes & $20$ nodes & $50$ nodes \\
    \midrule
    Sparse & $p=0.1^*$ & $m=1$ & $m=1$ & $m=2$ \\
    Dense & $p=0.4^*$ & $m=2$ & $m=4$ & $m=8$ \\
    \bottomrule
  \end{tabular}
    \begin{flushleft}
        \hspace{8.5em}\small{$^*$ Graphs are re-sampled such that they have at least $2$ edges.}
    \end{flushleft}
      \caption{Density schema for randomly sample graphs. The parameter $p$ denotes the probability of an edge between each pair of nodes in the graph, and $m$ denotes the average number of edges for each node in the graph. We scale the parameter $m$ with the number of nodes, such that the relative density (sparsity) is similar for all graph dimensions.}
\label{tab:density_er_sf}
\end{table}

\subsection{Additive noise model}\label{app:synthetic_data}
In this section, we provide a detailed description of the strategies for the generation of synthetic data under the additive noise model. 

\paragraph{Causal graph generation.} The simplest model for generation of causal DAG is the Erd\"{o}s-Renyi (ER) \citep{Erdos:1960}, which allows specifying the number of nodes $d$ and the average number of connections per node $m$ (or, alternatively, the probability $p$ of connecting each pair of nodes). In ER graphs, pairs of nodes have the same probability of being connected. Scale-free graphs (SF) are generated under a preferential attachment procedure \citep{Barabasi99emergenceScaling}, such that nodes with a higher degree are more likely to be connected with a new node, allowing for the presence of \textit{hubs} (i.e. high degree nodes) in the graphs. Scale-free properties are arguably characteristics of many real-world scenarios \citep{Barabasi99emergenceScaling}. In Table \ref{tab:density_er_sf} we report the schema defining the density of the edges relative to the number of nodes in the graph. Networks generated according to these two models are frequently used for evaluation of the performance of causal discovery algorithms (e.g. \citet{zheng18_notears, montagna23_das, montagna23_nogam, montagna23_assumption, rolland22_score, lachapelle19_grandag, ke23_csiva}).

\paragraph{Nonlinear causal mechanisms.} We consider two common practices for the generation of the nonlinear causal mechanisms of an additive noise model. We sample functions from a Gaussian process, such that $\forall i=1,\ldots,d$, $f_i(X_{PA_i}) = \mathcal{N}(\mathbf{0}, K(X_{PA_i}, X_{PA_i}))$, a multivariate normal distribution centered at zero and with covariance matrix as the Gaussian kernel $K(X_{PA_i}, X_{PA_i})$, where $X_{PA_i}$ are the observations of the parents of the node $X_i$. Another common approach is to define nonlinear mechanisms via neural networks, where the mechanism $f_i$ is defined as a multilayer perceptron (MLP) with a single hidden layer of $10$ nodes, \textit{leaky ReLU} nonlinear activation, and a normalizing layer. The weights of each network are initialized according to a standard normal distribution. These strategies for nonlinear mechanisms generation are commonly employed in previous works: a thorough list of references is provided in Section \ref{app:references} of the appendix.

 \paragraph{Additive noise distribution.} The additive noise terms $U_i$ are generated as nonlinear transformations $t : \R \rightarrow \R$ of a Gaussian random variable $N_i \hspace{.5mm} \mathsmaller{\sim} \hspace{.5mm} \mathcal{N}(0, \sigma_i)$, where $\sigma_i \hspace{.5mm} \mathsmaller{\sim} \hspace{.5mm} U(0.5, 1.0)$ uniformly distributed. In practice, for each node $i = 1, \ldots, d$, the corresponding noise term is defined as $U_i \coloneqq t(N_i)$, where $t$ is parametrized by an MLP with 100 nodes in the single hidden layer, sigmoid activation functions, and weights sampled from $U(-0.5, 0.5)$.

 \subsection{GP and NN data in the literature}\label{app:references}
 \textit{GP} and \textit{NN data} defined in this work are one of the most common ways (if not \textit{the} most common) to simulate nonlinear causal mechanisms for data generation, in order to evaluate empirical performance of causal discovery approaches. In what follows, we present a thorough list of papers in the causal discovery literature where nonlinear mechanisms are sampled from a gaussian process or a random neural network, similarly to our \textit{GP} and \textit{NN data}: \citet{mooij11_cyclic, buhlmann14_cam, peters14a_resit, mooij15_benchmark, louizos17_deeplatent, monti19_ica, lachapelle19_grandag, zhu20_reinforcement, brouillard20_differentiable, wang21_ordering, lippe2022efficient, rolland22_score, montagna23_assumption, chen2023_iscan, montagna23_das, reizinger23_jacobianbased, montagna23_nogam, ke23_csiva}.

\section{Proof of Proposition \ref{prop:identifiability}}\label{app:prop_proof}
\begin{proof}
    Consider the additive noise model $X \coloneqq U_X$, $Y \coloneqq f(U_X) + U_Y$. Given that the possible causal orderings of a bivariate graph are $(X, Y)$ and $(Y, X)$, showing that $\nu = 1$ is equivalent to proving that $\Var[s_X] > \Var[s_Y]$, where we define $s_X \coloneqq s_X(X, Y)$ and $s_Y \coloneqq s_Y(X, Y)$. By \eqref{eq:bivariate_score_x} and \eqref{eq:bivariate_score_y} we can derive the following expression of the score of $X$:
    \begin{equation}
            \Var[s_X] = \Varx + \Varf \Var[s_Y] + 2C, \label{eq:proof_scorex}
    \end{equation}
where $C \coloneqq \Covscore$. 

Then, from \eqref{eq:proof_scorex} we can rewrite the variance of $s_X$ as:
\begin{equation*}
    \Var[s_X] =  \Var[s_Y] \left(\Varf + \frac{\Varx}{\Vary} + \frac{2C}{\Vary} \right).
\end{equation*}
where, $\Var[s_X]$ is defined as $\Var[s_Y]$ multiplied by a coefficient. Let $\gamma \coloneqq \left(\Varf + \frac{\Varx}{\Vary} + \frac{2C}{\Vary} \right)$: it is immediate to see that for $\gamma > 1$, then $\Var[s_X] > \Var[s_Y]$, and vice-versa,  $\gamma \leq 1$ implies $\Var[s_X] \leq \Var[s_Y]$. Hence, we have that 
\begin{equation}
    \nu = 1 \Longleftrightarrow \left(\Varf + \frac{\Varx}{\Vary} + \frac{2C}{\Vary} \right) > 1.
    \label{eq:nu>1_gamma>1}
\end{equation}
By further manipulation of \eqref{eq:nu>1_gamma>1}, we obtain:
\begin{equation*}
    \nu = 1 \Longleftrightarrow \Varf > 1- \frac{\Varx}{\Vary} - \frac{2C}{\Vary}.
\end{equation*}
\end{proof}

\section{Score matching}\label{app:stein_gradient}
The goal of score matching is to infer the score function $s(x) \coloneqq
\nabla \log p(\mathbf{X})$ given an \textit{i.i.d.} sample $X = \{\mathbf{x}^{(k)}\}_{k=1,\ldots,n}$, extracted from the density $p$. In this section, we present a method developed in~\citet{stein_gradient} for estimating the score at the sample points, i.e., approximating $G \coloneqq (\nabla \log p(\mathbf{x}^1), \ldots, \nabla \log p(\mathbf{x}^n))^T \in \R^{n\times d}$. This resulting \textit{Stein gradient estimator} of the score is the one exploited by the ScoreSort algorithm (Algorithm \ref{alg:score-sort-population}). Our discussion will closely follow that of Section 2.2 of \citet{rolland22_score}.

This estimator is based on the Stein identity \citep{stein1972_identity}, which states that for any test function $\h:\R^d \rightarrow \R^{d'}$ such that $\lim_{\mathbf{x} \rightarrow \infty} \h(\mathbf{x}) p(\x) = 0$, we have
\begin{equation}\label{eq:3.2}
    \E_p[\h(\x) \nabla \log p(\x)^T + \nabla \h(\x)] = 0,
\end{equation}
where $\nabla \h(\x) \coloneqq (\nabla h_1(\x), \ldots, \nabla h_{d'}(\x))^T \in \R^{d' \times d}$.

By approximating the expectation in~\eqref{eq:3.2} using Monte Carlo, we obtain
\begin{equation}\label{eq:3.3}
    -\frac{1}{n} \sum_{k=1}^n \h(\x^{(k)}) \nabla \log p(\x^{(k)})^T + \text{err} = \frac{1}{n} \sum_{k=1}^n \nabla \h(\x^{(k)}),
\end{equation}
where $\text{err}$ is a random error term with mean zero, and which vanishes as $n \rightarrow \infty$ almost surely. By denoting $\textbf{H} = (\h(\x^{(1)}), \ldots, \h(\x^{(n)})) \in \R^{d'\times n}$ and $\overline{\nabla \h} = \frac{1}{n} \sum_{k=1}^n \nabla \h(\x^{(k)})$, equation~\eqref{eq:3.3} reads $-\frac{1}{n}\textbf{H} \textbf{G} + \text{err} = \overline{\nabla \h}$. Hence, by using ridge regression, the Stein gradient estimator is defined as:
\begin{equation}
\begin{split}
    \hat{\textbf{G}}^{\text{Stein}} &\coloneqq \argmin_{\hat{\textbf{G}}} \|\overline{\nabla \h} + \frac{1}{n}\textbf{H} \hat{\textbf{G}}\|_F^2 + \frac{\eta}{n^2} \|\hat{\textbf{G}}\|_F^2 \\
    &= -(\textbf{K} + \eta \textbf{I})^{-1} \langle \nabla, \textbf{K}\rangle,
\end{split}
    \label{eq:g_stein}
\end{equation}
where $\textbf{K} \coloneqq \textbf{H}^T\textbf{H}$, $\textbf{K}_{ij} = \kappa(\x^{(i)}, \x^{(j)}) \coloneqq \h(x^i)^T \h(\x^{(j)})$, $\langle \nabla, \textbf{K}\rangle = n \textbf{H}^T \overline{\nabla \h}$, $\langle \nabla, \textbf{K}\rangle_{ij} = \sum_{k=1}^n \nabla_{x_j^k} \kappa(\x^{(i)}, \x^{(k)})$ and $\eta \geq 0$ is a regularisation parameter. One can hence use the kernel trick, and use the estimator~\eqref{eq:g_stein} using any kernel $\kappa$ satisfying Stein's identity, such as the RBF kernel as shown in \citet{liu16_kernelized}.

\section{Hessian's estimator}\label{app:stein_hessian}
\citet{rolland22_score} extends score matching estimation of $\nabla \log p(\mathbf{X})$ by the Stein identity to the inference of the second order matrix of partial derivative $\nabla^2 \log p(\mathbf{X})$, Hessian of the log-likelihood. We propose an overview of the estimation procedure, closely following the discussion in Section 3.2 of \citet{rolland22_score}.

First, we need to introduce the second-order Stein identity \citep{diaconis2004_use, zhu21_hess}. Assuming that the distribution $p$ is twice differentiable, for any $q : \R^d \rightarrow \R$ such that $\lim_{\x \rightarrow \infty} q(\x) p(\x) = 0$ and such that $\E[\nabla^2 q(\x)]$ exists, the second-order Stein identity states that
\begin{equation}
    \E[q(\x) p(\x)^{-1} \nabla^2 p(\x)] = \E[\nabla^2 q(\x)],
\end{equation}
which can be rewritten as
\begin{equation}\label{eq:4.4}
\E[q(\x) \nabla^2 \log p(\x)] = \E[\nabla^2 q(\x) - q(\x) \nabla \log p(\x) \nabla \log p(\x)^T].
\end{equation}
In the case of SCORE, in order to identify a leaf of the causal graph we are only interested in estimating the diagonal elements of the score's Jacobian (the Hessian of the log-likelihood) at the sample points, i.e., $J \coloneqq (\text{diag}(\nabla^2 \log p(\x^{(1)})), \ldots, \text{diag}(\nabla^2 \log p(\x^{(n)})))^T \in \R^{n\times d}$. Using the diagonal part of the matrix equation~\eqref{eq:4.4} for various test functions gathered in $\h:\R^d \rightarrow \R^{d'}$, we can write
\begin{equation*}
\E[\h(\x) \text{diag}(\nabla^2 \log p(\x))^T] = \E[\nabla^2_{\text{diag}} \h(\x) - \h(\x) \text{diag}((\nabla \log p(\x) \nabla \log p(\x)^T))],
\end{equation*}
where $(\nabla^2_{\text{diag}} \h(\x))_{ij} = \frac{\partial^2 h_i(\x)}{\partial x_j^2}$. By approximating the expectations by an empirical average, we obtain, similarly as in~\eqref{eq:3.3},
%and the square operator in $(\nabla \log p(\x)^T)^2$ is applied component-wise
%
\begin{equation} \label{eq:4.6}
\frac{1}{n} \sum_{k=1}^n \h(\x^{(k)}) \text{diag}(\nabla^2 \log p(\x^{(k)}))^T + \text{err} = \frac{1}{n} \sum_{k=1}^n \nabla^2_{\text{diag}} \h(\x^{(k)}) - \h(\x^{(k)}) \text{diag}(\nabla \log p(\x^{(k)}) \nabla \log p(\x^{(k)})^T)).
\end{equation}

By denoting $\textbf{H} = (\h(\x^{(1)}), \ldots, \h(\x^{(n)})) \in \R^{d'\times n}$ and $\overline{\nabla^2_{\text{diag}} \h} \coloneqq \frac{1}{n} \sum_{k=1}^n \nabla^2_{\text{diag}} \h(\x^{(k)})$, equation~\eqref{eq:4.6} reads $\frac{1}{n} \textbf{H} \textbf{J} + \text{err} = \overline{\nabla^2_{\text{diag}} \h} - \frac{1}{n} \textbf{H} \text{diag}(\textbf{G}\textbf{G}^T)$. Hence, by using the Stein gradient estimator for $\textbf{G}$, we define the Stein Hessian estimator as the ridge regression solution of the previous equation, i.e.
\begin{equation}
\begin{split}
    &\hat{\textbf{J}}^{\text{Stein}} \coloneqq \argmin_{\hat{\textbf{J}}} \left \|\frac{1}{n} \textbf{H} \hat{\textbf{J}} + \frac{1}{n} \textbf{H} \text{diag}\left(\hat{\textbf{G}}^{\text{Stein}}\left(\hat{\textbf{G}}^{\text{Stein}}\right)^T\right) - \overline{\nabla^2_{\text{diag}} \h} \right\|_F^2 + \frac{\eta}{n^2} \|\hat{\textbf{J}}\|_F^2 \\
    &= -\text{diag}\left(\hat{\textbf{G}}^{\text{Stein}}\left(\hat{\textbf{G}}^{\text{Stein}}\right)^T\right) + (\textbf{K} + \eta \textbf{I})^{-1} \langle \nabla^2_{\text{diag}}, \textbf{K}\rangle, 
\end{split}
    \label{eq:j_stein}
\end{equation}

where $\textbf{K}_{ij} = \kappa(\x^{(i)}, \x^{(j)}) \coloneqq \h(\x^{(i)})^T \h(\x^{(j)})$, $\langle \nabla^2_{\text{diag}}, \textbf{K}\rangle = n \textbf{H}^T \overline{\nabla^2_{\text{diag}} \h}$, $\langle \nabla^2_{\text{diag}}, \textbf{K}\rangle_{ij} = \sum_{i=1}^n \frac{\partial^2 \kappa(\x^{(i)}, \x^{(k)})}{\partial (\x^{(k)}_j)^2}$ and $\textbf{G}^{\text{Stein}}$ is defined in~\eqref{eq:g_stein}.

\section{Proof of Proposition \ref{prop:stat_efficiency}}\label{app:stat_efficiency}
\begin{proof}
    Let $G \coloneqq (\nabla \log p(\mathbf{x}^1), \ldots, \nabla \log p(\mathbf{x}^n))^T \in \R^{n\times d}$ be the matrix of the score for each observation $\x^{(k)}$ of the sample. Let $\hat{G} \coloneqq \hat{G}^{\textnormal{stein}}$ defined in \eqref{eq:g_stein}.
We define the matrix of the estimation errors as:
\begin{equation}
\begin{split}
    &\Delta_{G} \coloneqq G - \hat{G} \\ &= (\partial_{x_i} \log p(\x^{(k)}) - \partial_{x_i} \widehat{\log p(\x^{(k)})})_{1 \leq k \leq n, \hspace{1mm} 1 \leq i \leq d} \in \R^{n \times d}.
\end{split}
\label{eq:delta_g}
\end{equation}

Similarly, we define the estimation error on the diagonal terms of the score's Jacobian. Let $\hat{J}(\hat{G}) \coloneqq \hat{J}^{\textnormal{stein}}$ defined in \eqref{eq:j_stein}, where the argument $\hat{G}$ is used to remark the dependence from the Stein gradient estimator of \eqref{eq:g_stein}.
The resulting matrix of statistical errors is :
\begin{equation}
\begin{split}
    &\Delta_{J(\hat{G})} \coloneqq J - \hat{J}(\hat{G}) \\ &= (\partial^2_{x_i} \log p(\x^{(k)}) - \partial^2_{x_i} \widehat{\log p(\x^{(k)})})_{1 \leq k \leq n, \hspace{1mm} 1 \leq i \leq d} \in \R^{n \times d}.
\end{split}
\end{equation}

Now, consider the case where $\hat{G} = G$, i.e. we have perfect estimates of the score: from \eqref{eq:j_stein}, we have that the score's Jacobian optimal estimator is:
\begin{equation}
\begin{split}
    \hat{J}(G) \coloneqq -\text{diag}\left(GG^T\right) + (K + \eta \textbf{I})^{-1} \langle \nabla^2_{\text{diag}}, K\rangle, 
\end{split}
    \label{eq:j_stein_g}
\end{equation}
which, by assumption, is subject to zero error, i.e. $J - \hat{J}(G) = 0$.

\paragraph{Remark.} We defined the matrix $\Delta_G$ of error in the estimation of the score $\nabla \log p(\mathbf{X})$ by using \eqref{eq:g_stein}. Similarly, we define $\Delta_{J(\hat{G})}$  error of estimation when the Hessian of the log-likelihood is computed by \eqref{eq:j_stein}, as a function of $\hat{G}$. In the case where the matrix $G$ is exactly known, then the statistical error in the inference of $J$ is null by hypothesis.

We want to show that errors in the estimation of $G$ propagate to $\hat{J}(\hat{G})$. We start manipulating the expression of $\hat{J}(\hat{G})$ of \eqref{eq:j_stein}:
\begin{equation}
\begin{split}
&\hat{J}(\hat{G}) = -\text{diag} \left(\hat{G}\hat{G}^T \right) + (\textbf{K} + \eta \textbf{I})^{-1} \langle \nabla^2_{\text{diag}}, \textbf{K}\rangle \\
&= -\text{diag} \left(\hat{G}\hat{G}^T + GG^T - GG^T \right) + (\textbf{K} + \eta \textbf{I})^{-1} \langle \nabla^2_{\text{diag}}, \textbf{K}\rangle \\
&= -\text{diag} \left(GG^T \right) + (\textbf{K} + \eta \textbf{I})^{-1} \langle \nabla^2_{\text{diag}}, \textbf{K}\rangle + \textnormal{diag}\left( GG^T - \hat{G}\hat{G}^T \right) \\
&= \hat{J}(G) + \textnormal{diag}\left( (G-\hat{G})(G+\hat{G})^T \right) \\
&= \hat{J}(G) + \textnormal{diag}\left( \Delta_G(G+\hat{G})^T \right) \\
\end{split}
\end{equation}
Hence, we see that $\hat{J}(\hat{G})$ is equivalent to the ridge regression solution of \eqref{eq:j_stein_g}, namely the score's Jacobian estimator computed with the exact value of $G$, plus an error term that propagates from the first order estimates of the score. Given the assumption $J - \hat{J}(G) = 0$, we get:
\begin{equation*}
    \begin{split}
        \abs*{J - \hat{J}(\hat{G})} &= \abs*{J - \hat{J}(G) + \textnormal{diag}\left( \Delta_G(G+\hat{G})^T \right)} \\
        &= \abs*{\Delta_G(G+\hat{G})^T},
    \end{split}
\end{equation*}
such that for each sample $\mathbf{x}^{(k)}$ in the dataset, the resulting error is defined as:
\begin{equation*}
    \begin{split}
        \epsilon_i^{(k)} &\coloneqq \abs*{\partial^2_{x_i} \log p(\mathbf{x}^{(k)}) - \widehat{\partial^2_{x_i} \log p(\mathbf{x}^{(k)})}}\\
        &= \abs*{\partial_{x_i} \log p(\mathbf{x}^{(k)}) - \widehat{\partial_{x_i} \log p(\mathbf{x}^{(k)})}} \abs*{\partial_{x_i} \log p(\mathbf{x}^{(k)}) + \widehat{\partial_{x_i} \log p(\mathbf{x}^{(k)})}}.
    \end{split}
\end{equation*}
\end{proof}

\section{Other patterns of sortability in the nonlinear additive noise model}
In this section, we provide an overview of the \textit{varsortability} and \textit{R$^2$-sortability}, two patterns emerging in the setting of linear SCM which can inform about the causal order of the model by simple sorting heuristics.

\subsection{Varsortability}
\citet{reisach21_beware} shows that when the causal mechanisms of the model generating the data are linear, it is possible to identify the topological order of the variable by simple sorting of the variables by ascending order of their variance. In particular, assuming the bivariate causal relation $Y = wX + N_Y$, where $w$ is the linear coefficient of the structural equation, identifiability of the order by the variance of $X$ and $Y$ is verified if and only if $w^2\Var[X] + \Var[N_Y] > \Var[X]$, which is equivalent to $\Var[Y] > \Var[X]$. Under the hypothesis of a nonlinear causal model $Y = f(X) + N_Y$, the condition on the variance becomes $\Var[f(X)] + N_Y > \Var[X]$. The experiments in \citet{reisach21_beware} show that \textit{varsortability} is a common feature of simulated additive noise model data, in the case of both linear and nonlinear mechanisms.

\subsection{R$^2$-sortability.} 
Closely related to the varsortability of causal models, \citet{reisach2023_simple} recently identified another pattern emerging in synthetic data generated under the linear model $$\mathbf{X} = W\mathbf{X} + \mathbf{N},$$ where $W$ denotes the weight matrix and $\mathbf{N}$ is the random vector of the noise terms. The marginal variance of a random variable $X_i$ is defined by $\Var[X_i] = \Var[\mathbf{W}_i]$
Given that varsortability is implied by the increasing marginal variance $\Var[X_i] = \Var[\mathbf{W}_i^T\mathbf{X}] + \Var[N_i]$, with $\mathbf{W}_i$ denoting the $i$-th row of the matrix W, the intuition is that the variance $\Var[\mathbf{W}_i^T\mathbf{X}]$ explained by the parents of a node also increases in the causal direction. Then, the vector defined by cause-explained variance fraction $(\frac{\Var[\mathbf{W}_i^T\mathbf{X}]}{\Var[X_i]})_i$ may provide information about the causal ordering of the model. Given that the cause-explained variance of a variable can not be directly estimated from the data, the authors of the paper define the coefficient $R^2_i \coloneqq 1 - \frac{\Var[X_i - \Mu[X_i | \mathbf{X}\setminus \{X_i\}]]}{\Var[X_i]}$ as an upper bound of the cause-explained variance, where the expectation $\Mu[X_i | \mathbf{X}\setminus \{X_i\}]$ can be inferred by regressing $X_i$ on all the remaining nodes in the graph. Then, a model is said to be R$^2$-sortable when the causal order is found by sorting the vector of $(R^2_i)_i$ coefficients by their ascending value. 

\section{Experiments on Scale-free graphs}\label{app:scale-free-exp}
\begin{figure}
    \centering
    \begin{subfigure}[b]{1\textwidth}
        \centering
        \includegraphics[width=0.57\textwidth]{Figures/legend_scoresort.png}
    \end{subfigure}
     \begin{subfigure}[b]{1\textwidth}
        \centering        
    \includegraphics[width=0.8\textwidth]{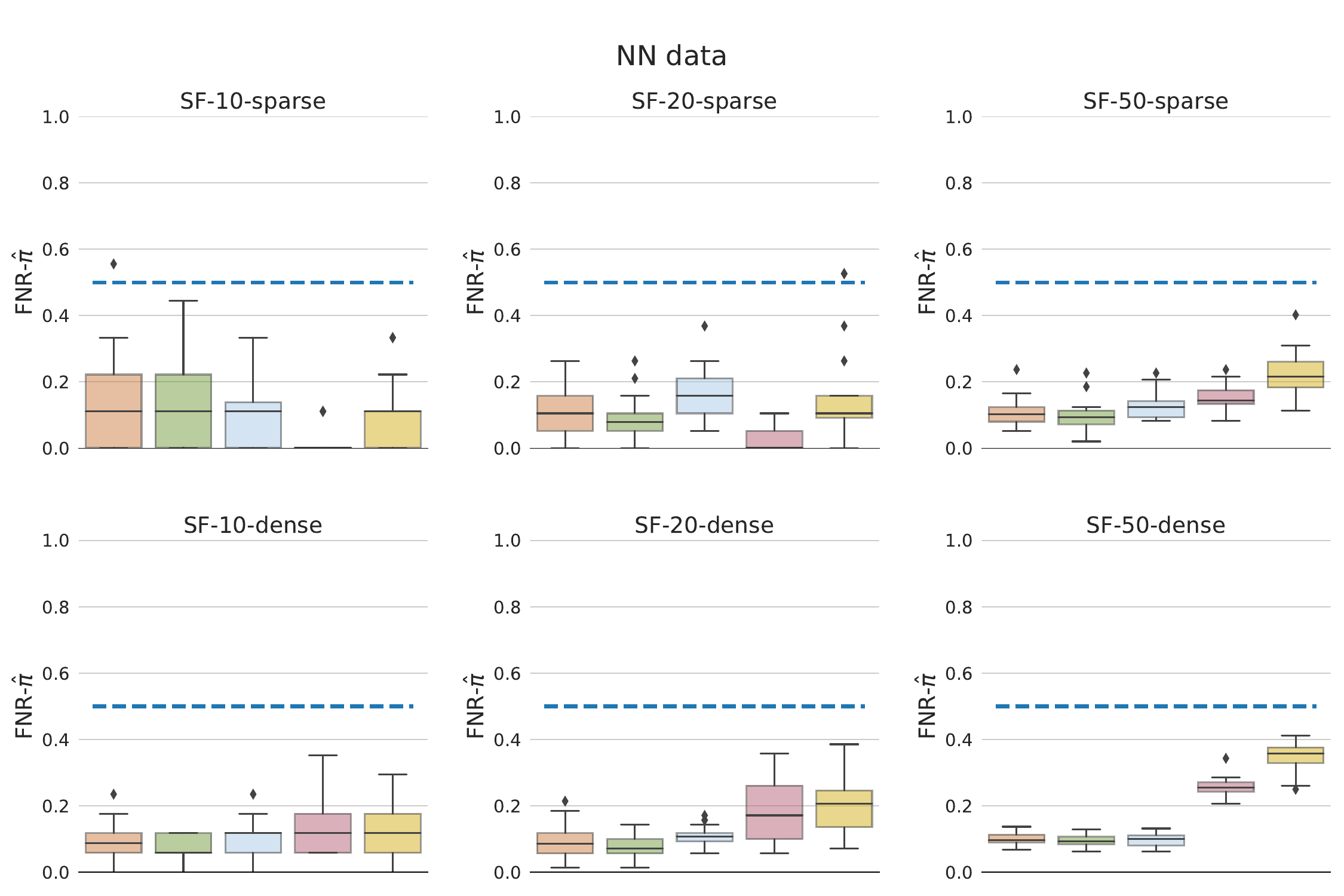}
    \end{subfigure}
    \caption{\footnotesize{Experimental results on dense and sparse SF with number of nodes in the set $\{10, 20, 50\}$ and GP causal mechanisms. The graphs show the $\textnormal{FNR-}\hat{\pi}$ accuracy (the lower, the better) of the order estimates, with the boxplots evaluated over $20$ random seeds. Score-sortability is defined as $\nu = 1 - \textnormal{FNR-}\hat{\pi}$ achieved by ScoreSort, such that low values of $\textnormal{FNR-}\hat{\pi}$ denotes high score-sortability of the model. The dashed blue line is the expected accuracy of random ordering.}}
    \label{fig:exp_GP_SF}
\end{figure}

\begin{figure}
    \centering
    \begin{subfigure}[b]{1\textwidth}
        \centering
        \includegraphics[width=0.57\textwidth]{Figures/legend_scoresort.png}
    \end{subfigure}
     \begin{subfigure}[b]{1\textwidth}
        \centering        
    \includegraphics[width=0.8\textwidth]{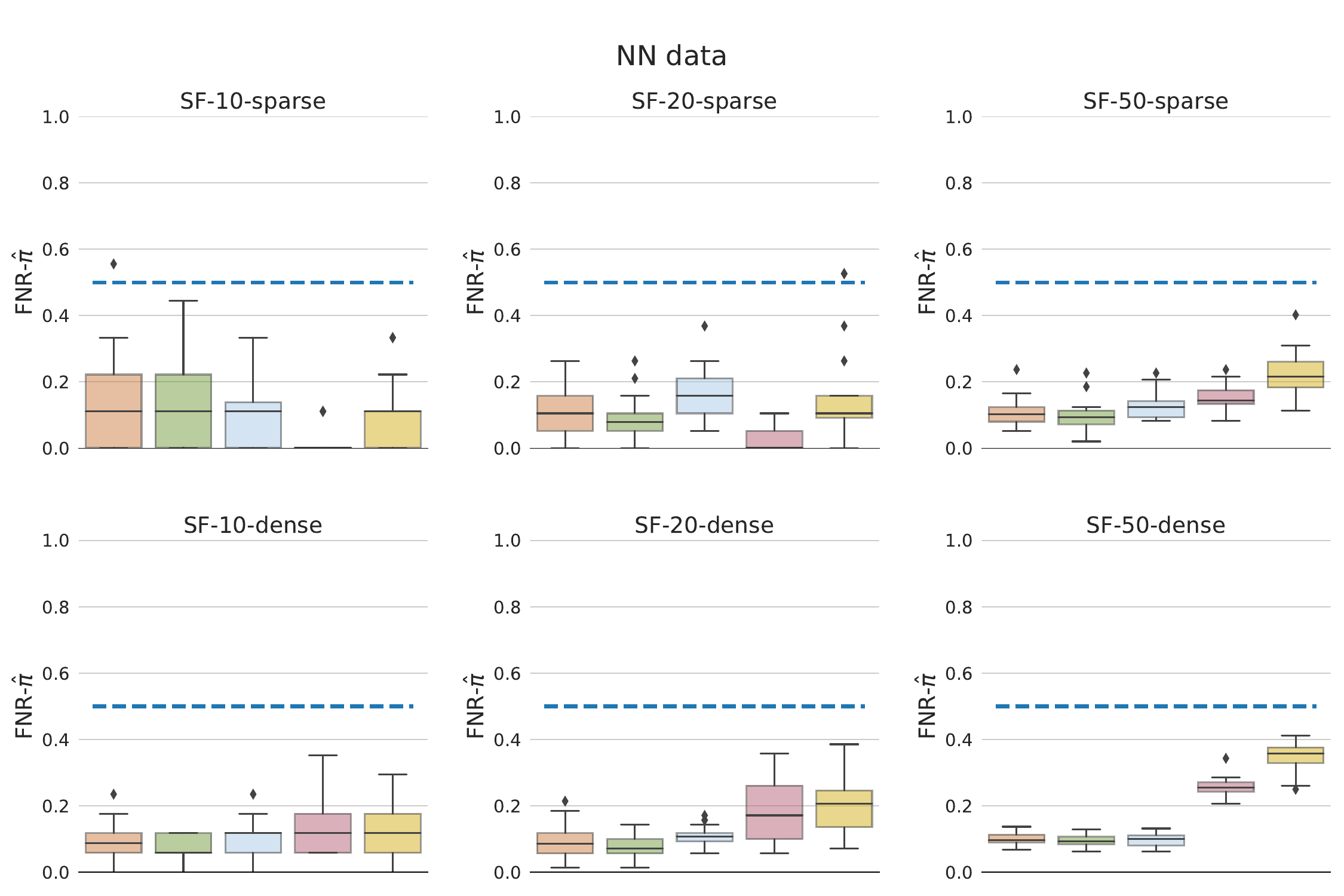}
    \end{subfigure}
    \caption{\footnotesize{Experimental results on dense and sparse SF with number of nodes in the set $\{10, 20, 50\}$ and NN causal mechanisms. The graphs show the $\textnormal{FNR-}\hat{\pi}$ accuracy (the lower, the better) of the order estimates, with the boxplots evaluated over $20$ random seeds. Score-sortability is defined as $\nu = 1 - \textnormal{FNR-}\hat{\pi}$ achieved by ScoreSort, such that low values of $\textnormal{FNR-}\hat{\pi}$ denotes high score-sortability of the model. The dashed blue line is the expected accuracy of random ordering.}}
    \label{fig:exp_NN_SF}
\end{figure}

\begin{figure}
    \centering
    \begin{subfigure}[b]{1\textwidth}
        \centering
        \includegraphics[width=0.57\textwidth]{Figures/legend_scoresort.png}
    \end{subfigure}
     \begin{subfigure}[b]{1\textwidth}
        \centering        
    \includegraphics[width=.8\textwidth]{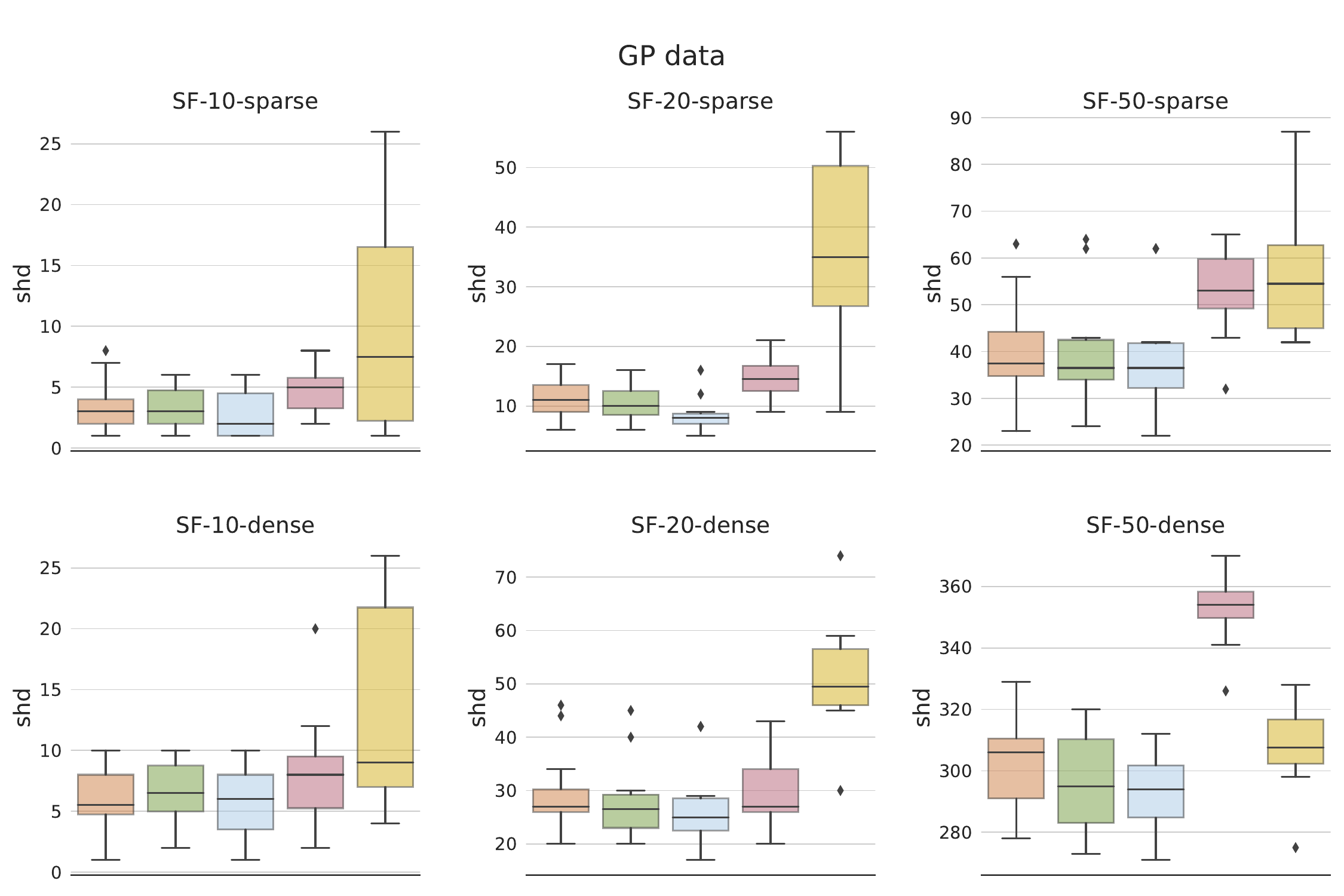}
    \end{subfigure}
    \caption{\footnotesize{Structural Hamming Distance (the lower, the better) on dense and sparse SF with the number of nodes in the set $\{10, 20, 50\}$ and GP causal mechanisms. Box plots are evaluated over $20$ random seeds.}}
    \label{fig:exp_GP_SF_shd}
\end{figure}

\begin{figure}
    \centering
    \begin{subfigure}[b]{1\textwidth}
        \centering
        \includegraphics[width=0.57\textwidth]{Figures/legend_scoresort.png}
    \end{subfigure}
     \begin{subfigure}[b]{1\textwidth}
        \centering        
    \includegraphics[width=.8\textwidth]{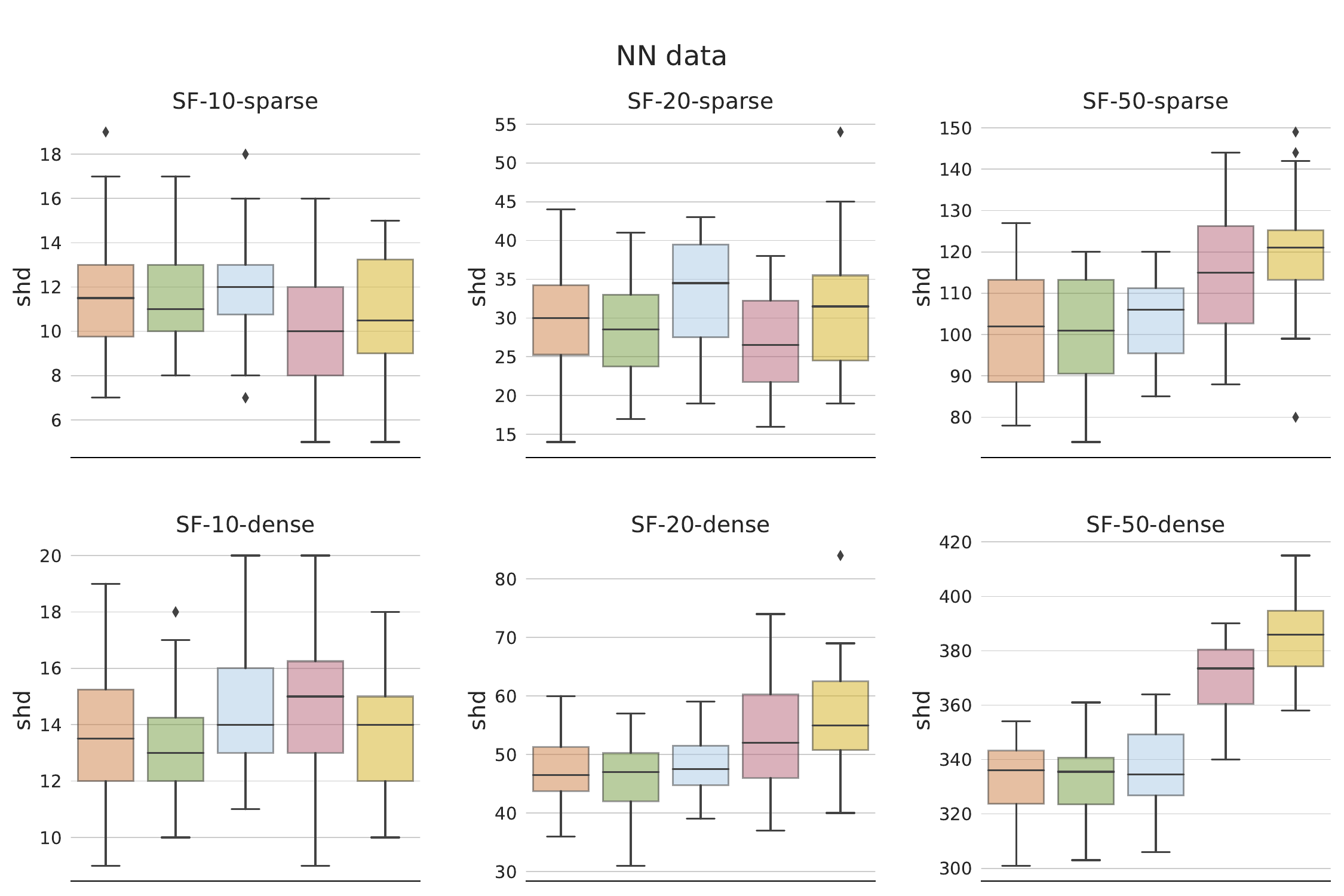}
    \end{subfigure}
    \caption{\footnotesize{Structural Hamming Distance (the lower, the better) on dense and sparse SF with the number of nodes in the set $\{10, 20, 50\}$ and NN causal mechanisms. Box plots are evaluated over $20$ random seeds.}}
    \label{fig:exp_NN_SF_shd}
\end{figure}

Figure \ref{fig:exp_GP_SF} and \ref{fig:exp_NN_SF} show the $\textnormal{FNR-}\hat{\pi}$ of ScoreSort, SCORE, NoGAM, CAM, and RESIT on sparse and dense Scale-free graphs with $\{10, 20, 50\}$ nodes, and causal mechanisms sampled from Gaussian processes (\textit{GP data}) and random neural networks (\textit{NN data}). We observe that similarly to the case of Erd\"{o}s-Renyi causal graphs, these common benchmarks tend to be score-sortable. Additionally, we report the Structural Hamming Distance on Scale-free graphs in Figure \ref{fig:exp_GP_SF_shd} for \textit{GP data} and Figure \ref{fig:exp_NN_SF_shd} for \textit{NN data}.

% SF FNR-pi

% SF NN FNR-pi

% SF GP SHD

% SF NN SHD

% -------------------------------
\section{Structural hamming distance on Erd\"{o}s-Renyi graphs} \label{app:er_shd}
\begin{figure}
    \centering
    \begin{subfigure}[b]{1\textwidth}
        \centering
        \includegraphics[width=0.57\textwidth]{Figures/legend_scoresort.png}
    \end{subfigure}
     \begin{subfigure}[b]{1\textwidth}
        \centering        
    \includegraphics[width=.95\textwidth]{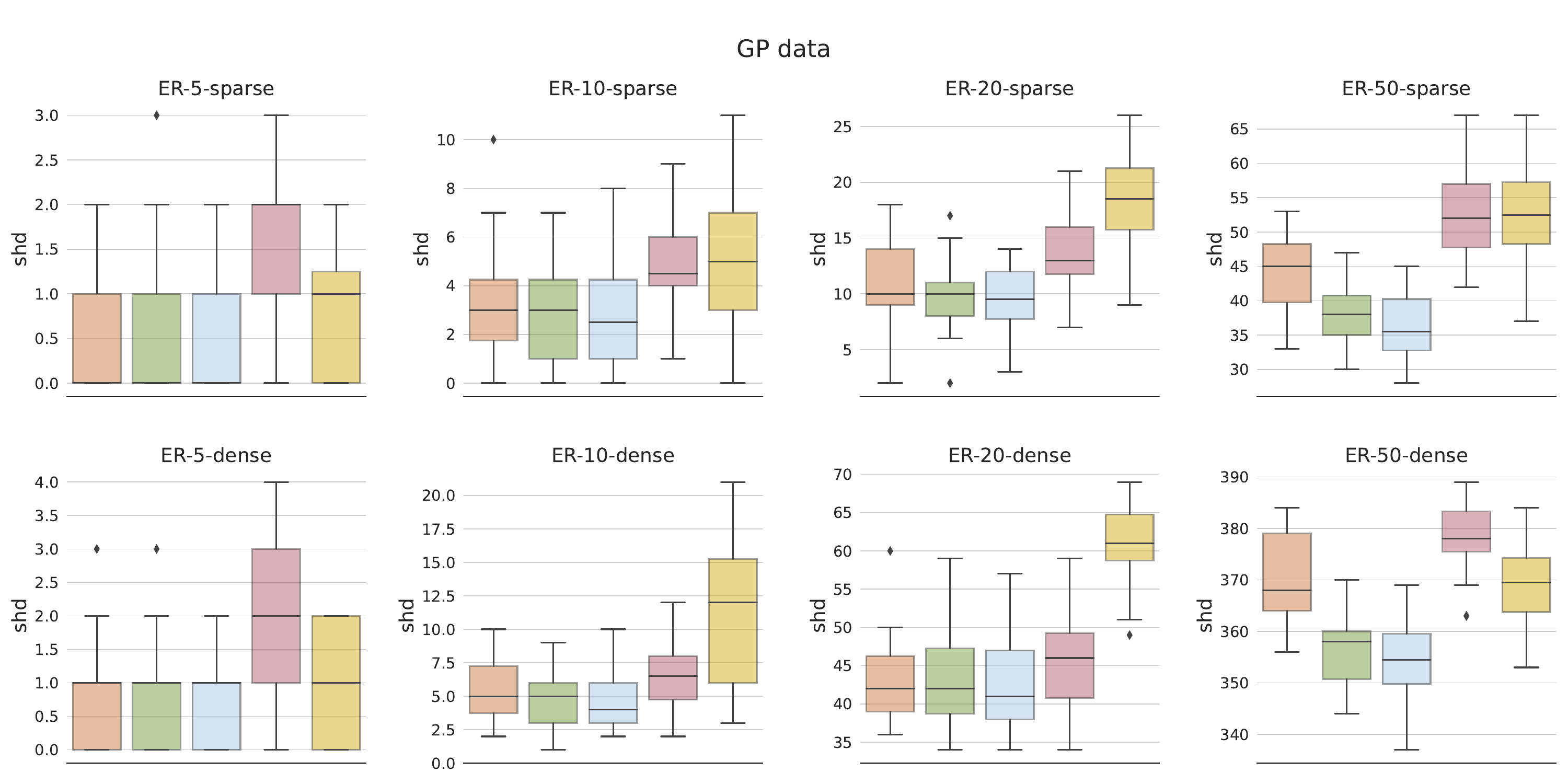}
    \end{subfigure}
    \caption{\footnotesize{Structural Hamming Distance (the lower, the better) on dense and sparse ER with the number of nodes in the set $\{5, 10, 20, 50\}$ and GP causal mechanisms. Box plots are evaluated over $20$ random seeds.}}
    \label{fig:exp_GP_ER_shd}
\end{figure}

\begin{figure}
    \centering
    \begin{subfigure}[b]{1\textwidth}
        \centering
        \includegraphics[width=0.57\textwidth]{Figures/legend_scoresort.png}
    \end{subfigure}
     \begin{subfigure}[b]{1\textwidth}
        \centering        
    \includegraphics[width=.95\textwidth]{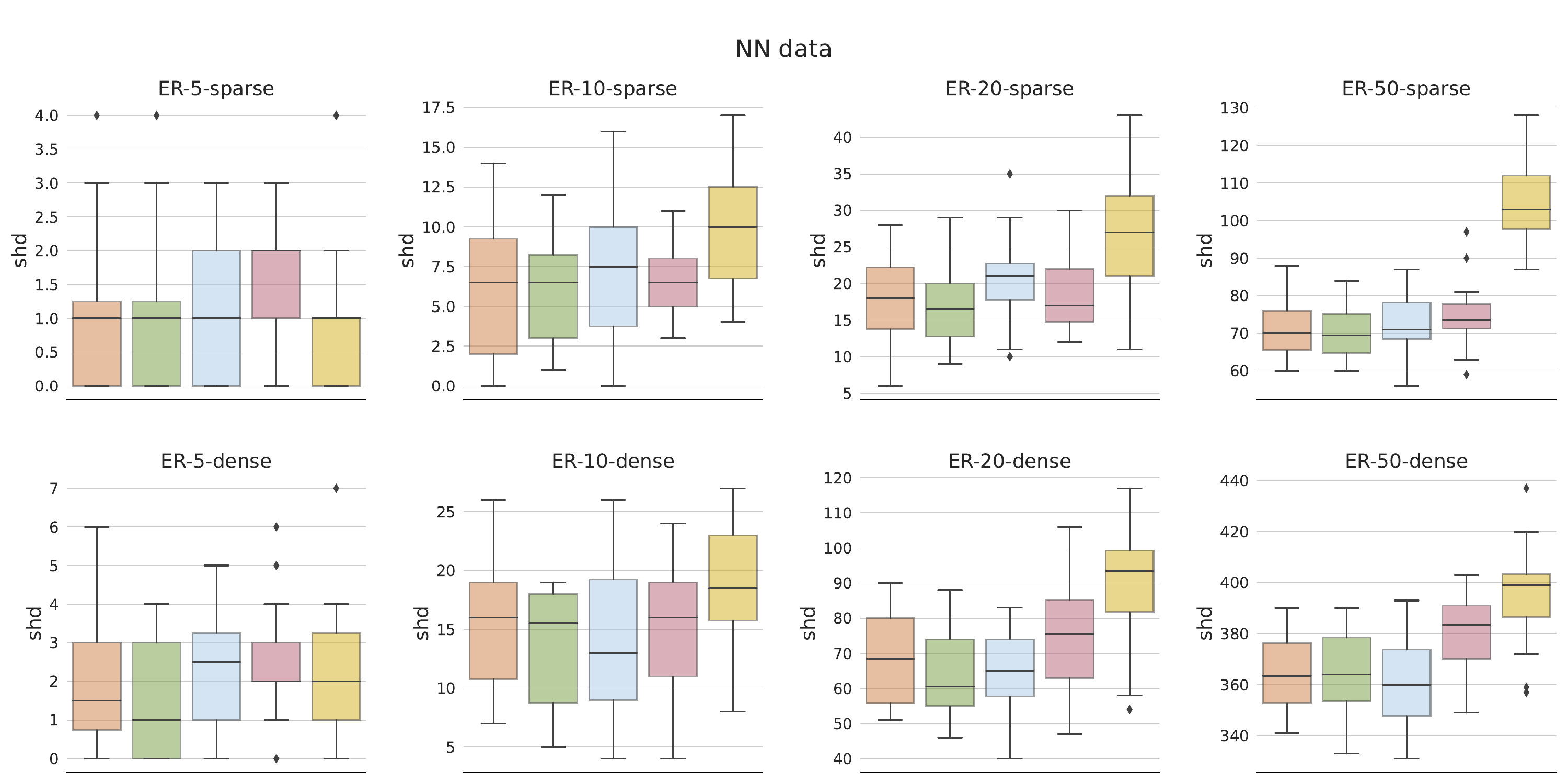}
    \end{subfigure}
    \caption{\footnotesize{Structural Hamming Distance (the lower, the better) on dense and sparse ER with the number of nodes in the set $\{5, 10, 20, 50\}$ and NN causal mechanisms. Box plots are evaluated over $20$ random seeds.}}
    \label{fig:exp_NN_ER_shd}
\end{figure}

In this section, we report the SHD of the experiments discussed in Section \ref{sec:experiments} of the main manuscript. Figure \ref{fig:exp_GP_ER_shd} shows results for \textit{GP data}, whereas \ref{fig:exp_NN_ER_shd} refers to the inference on \textit{NN data}. All the benchmarked methods perform the edge selection step via CAM-pruning procedure (see Appendix \ref{app:cam}), with $\alpha=0.05$ for the \textit{p-value} thresholding.

% ER GP SHD

% ER NN SHD

\end{document}